\documentclass[a4paper,11pt]{article}
\usepackage{jinstpub} 
\usepackage{graphicx}[pdftex]
\usepackage{booktabs}
\usepackage{hyperref}
\usepackage{siunitx}
\usepackage{threeparttable}
\usepackage{subfigure}
\usepackage{changepage}
\usepackage{textgreek}
\usepackage{verbatim}
\usepackage{multirow}
\usepackage{tabularx}
\usepackage[scaled=.90]{helvet} 
\usepackage{multirow}
\usepackage{courier}
\usepackage[dvipsnames,svgnames]{xcolor}
\usepackage{soul}
\usepackage[version=4]{mhchem}
\usepackage{hyperref}

\usepackage{natbib}
\newcommand\BaF {BaF$_2$ }
\newcommand\BaFY {BaF$_2$:Y }

\title{Exploring Scintillators and Cherenkov Radiators for MIP Timing Detectors}

\author[a,b]{R.~Cala'}
\author[a,b]{L.~Martinazzoli$^{*,}$}
\author[a]{N.~Kratochwil\footnote{now at Department of Biomedical Engineering, College of Engineering, University of California, Davis}}
\author[a,c]{I.~Frank}
\author[a,b]{M.~Salomoni}
\author[a,b]{F.~Pagano\footnote{now at Institute of Instrumentation for Molecular Imaging (I3M), Valencia, Spain}}
\author[a,b]{G.~Terragni}
\author[a,d]{C.~Lowis}
\author[e]{J.~Chen}
\author[f]{J.~Pejchal}
\author[f]{P.~Bohacek}
\author[f]{M.~Nikl}
\author[g]{S.~Tkachenko}
\author[g]{O.~Sidlestkiy}
\author[b]{M.~Paganoni}
\author[a,b]{M.~Pizzichemi}
\author[a]{E.~Auffray}
\affiliation[a]{European Organization for Nuclear Research (CERN), Geneva, Switzerland}
\affiliation[b]{INFN \& Università degli Studi di Milano-Bicocca, Milan, Italy}
\affiliation[c]{Ludwig Maximilian University, Munich, Germany}
\affiliation[d]{RWTH Aachen University, Aachen, Germany}
\affiliation[e]{Shanghai Institute of Ceramics, Chinese Academy of Sciences, Shanghai 201899, PR China}
\affiliation[f]{Institute of Physics of the Czech Academy of Sciences, Prague, Czech Republic}
\affiliation[g]{Institute for Scintillation Materials NAS of Ukraine, Kharkiv, Ukraine}

\emailAdd{loris.martinazzoli@cern.ch}

\abstract{
This article presents the timing performance of materials with fast light emission, tested as Minimum Ionizing Particle detectors using 150 GeV hadron beams in Monte Carlo simulations and at the CERN SPS North Area. Pixels of cross-section 2\,$\times$\,2\,mm$^2$ or 3\,$\times$\,3\,mm$^2$ and length of 3 or 10~mm were coupled to Hamamatsu SiPM and read out by fast high-frequency electronics.
Materials whose timing performance relies on Cherenkov emission, namely BGSO, PWO, and PbF$_2$, achieved time resolutions in the range 24-36\,ps.
Scintillators as L(Y)SO:Ce, GAGG, and \BaF reached below 15\,ps, the best topping at 12.1\,$\pm$\,0.4\,ps.
These fast materials are compared to LYSO and their additional benefit is discussed.
Given the promising results of \ce{BaF2}, the study is completed with measurements of the scintillation properties of a set doped with yttrium to quench the slow light emission.

}

\keywords{Calorimeters, Cherenkov detectors, Scintillators, scintillation and light emission processes (solid, gas and liquid scintillators), Timing detectors}

\arxivnumber{2411.06977} 

\begin{document}
\maketitle
\flushbottom

\section{Introduction}

Timing is going to be crucial for particle detection at future high energy physics experiments~\cite{ECFA_2021, FCC:2018vvp}. Precise physics measurements require larger datasets and colliders must run at higher instantaneous luminosity to provide them. The consequent density of tracks and pile-up of events will pose a difficult challenge to the reconstruction algorithms, which will suffer from combinatorial background. Time acts as an additional dimension to mitigate the challenges of increased luminosity and disentangle the multitude of hits in the subdetectors, assigning them to the correct track.
The precision required is normally a few tens of picoseconds~\cite{CMS, CERN-LHCC-2017-011, LHCB_FTDR, AleksaFCCee}; therefore, the community is exploring solutions to integrate the time information into future detectors, be it in the calorimeter~\cite{spacal_2023} or through \emph{ad~hoc} timing layer for minimum ionizing particles (MIP)~\cite{CMS_TL}. 
This article focuses on the latter.

R\&D is ongoing to investigate new materials and processes capable of fulfilling the demand for fast detectors both in medical imaging and high energy physics applications~\cite{Lecoq_2020}. The possibility to approach a precision of a few tens of picoseconds was already demonstrated using fast inorganic scintillators -- e.g. L(Y)SO and aluminium garnet crystals -- coupled to silicon photomultiplier devices (SiPMs)~\cite{Benaglia-2016, Lucchini-2017, Abbott_2021}.
This work studies the timing capabilities of attractive novel materials coupled to SiPMs under 150~GeV pion irradiation at SPS proton accelerator facility at CERN. 
Sec.~\ref{sec:TB} and \ref{sec:MC} investigate multiple light-emission processes via testbeam and Monte Carlo simulations, including scintillation, Cherenkov emission, and cross-luminescence. Fast amplifiers based on state-of-the-art high-frequency electronics were employed~\cite{Gundacker-2019}. Moreover, we explore the effect of time-walk for different materials and the relation between time resolution and scintillation characteristics.

Finally, given the promising performance of the cross-luminescent \BaF, we dedicate Sec.~\ref{sec:baf2} to the laboratory characterisation of a set of \ce{BaF2} with various yttrium doping, measuring their time and scintillation characteristics.

\section{Materials}
The samples tested are listed in Table~\ref{tab:List_samples_MIPs} and their scintillating properties in Table\,\ref{tab:Table2}. The samples included lutetium oxyorthosilicates (L(Y)SO), gadolinium aluminium garnets (GAGG), bismuth oxides (B(GS)O), lead tungstate PbWO$_4$ (PWO) and fluoride (PbF$_2$), barium fluorides (BaF$_2$) and a plastic scintillator (EJ232). The two samples of GAGG Accelerated were heavily co-doped with Cerium and Magnesium to accelerate the scintillation beyond current commercial values~\cite{Martinazzoli-2022}.
All presented mirror-polished surfaces and no apparent defects, but for the BGSO and the FZU GAGG crystal from the first production batch which showed some cracks or inclusions.
The samples were wrapped in several Teflon layers and coupled with optical glue (Cargille Meltmount) to S13360-3050PE SiPMs (Hamamatsu Photonics, 3\,$\times$\,3~mm$^2$ active area, 50~$\mu$m SPAD size) -- except for the \ce{BaF2} and \ce{BaF2}:Y due to the deep UV emission spectrum.

\ce{BaF2} were wrapped in several Teflon layers and coupled to SiPMs with Viscasil 100M optical grease, selected thanks to the transparency to the cross-luminescent emission of \ce{BaF2} in the deep UV \cite{Klamra-1987, Pots-2020}.
\ce{BaF2} 3\,$\times$\,3\,$\times$\,10~mm$^3$ samples were coupled to S13370-6050CN SiPMs (Hamamatsu Photonics, 6\,$\times$\,6~mm$^2$ active area, 50 $\mu$m SPAD size).  \ce{BaF2} 2\,$\times$\,2\,$\times$\,10~mm$^3$ samples were coupled to S13370-3050CN SiPMs (Hamamatsu Photonics, 3\,$\times$\,3~mm$^2$ active area, 50 $\mu$m SPAD size).These SiPMs were biased at 58~V.
Lacking glue's mechanical stability, \ce{BaF2} samples were held in position by a 3D-printed plastic holder.

\begin{table}[] {\small 
    \centering
    \caption{List of the samples measured during the two testbeams, including their producer and size. The checkmarks indicate whether the sample was measured in the 2022, 2023, or both testbeam campaigns.}
    \label{tab:List_samples_MIPs}
    \begin{tabularx}{\textwidth}{@{\extracolsep{\fill}}lllcc}
        \toprule
           
        \textbf{Sample}   & \textbf{Producer}    & \textbf{Dimensions} & \multicolumn{2}{l}{\textbf{Testbeam campaign}} \rule{0pt}{3ex}\\
        \textbf{material} &                       & [mm$^3$]           & \textbf{2022} & \textbf{2023}   \rule[-1.5ex]{0pt}{0ex} \\       
        \midrule
        
        LYSO:Ce           & \textit{CPI }               &   2\,$\times$\,2\,$\times$\,10  & \checkmark &   \rule{0pt}{2.5ex} \\
        LSO:Ce,Ca         & \textit{Agile}              &   2\,$\times$\,2\,$\times$\,10  & \checkmark &  \rule{0pt}{2.5ex}\\
        GFAG          & \textit{C\&A}                &   2\,$\times$\,2\,$\times$\,10  & \checkmark & \checkmark \rule{0pt}{5.0ex}\\
        GAGG:Ce:Mg Accelerated (1$^\text{st}$ batch)  & \textit{FZU Prague}          &   2\,$\times$\,2\,$\times$\,10  & \checkmark &    \rule{0pt}{2.5ex}\\
        GAGG:Ce:Mg Accelerated (2$^\text{nd}$ batch)  & \textit{FZU Prague}          &   2\,$\times$\,2\,$\times$\,10  & & \checkmark   \rule{0pt}{2.5ex}\\
        EJ232                      & \textit{Eljen Technology}    &   3\,$\times$\,3\,$\times$\,3  & \checkmark  & \rule{0pt}{5.0ex}\\
        BGO             & \textit{EPIC}                &   2\,$\times$\,2\,$\times$\,10  & \checkmark  & \checkmark   \rule{0pt}{5.0ex}\\
        BGSO                  & \textit{ISMA}                &   2\,$\times$\,2\,$\times$\,10  & \checkmark  & \checkmark \rule{0pt}{2.5ex}\\
        BSO                   & \textit{SICCAS}                &   2\,$\times$\,2\,$\times$\,10 &  & \checkmark  \rule{0pt}{2.5ex}\\
        PWO             &                              &   2\,$\times$\,2\,$\times$\,10 & \checkmark  &    \rule{0pt}{2.5ex}\\
        PbF$_2$          &  \textit{EPIC}               &   2\,$\times$\,2\,$\times$\,10  & \checkmark  &  \rule{0pt}{2.5ex}\\
        BaF$_2$           &  \textit{SICCAS}             &   3\,$\times$\,3\,$\times$\,10  & \checkmark  &  \rule{0pt}{5.0ex}\\
        BaF$_2$:\,1\,mol.\%\,Y         &  \textit{SICCAS}   &   3\,$\times$\,3\,$\times$\,10  & \checkmark  & \rule{0pt}{2.5ex}\\
        BaF$_2$:\,3\,mol.\%\,Y          &  \textit{SICCAS}   &   3\,$\times$\,3\,$\times$\,10  & \checkmark  &  \rule{0pt}{2.5ex}\\
        BaF$_2$:\,5\,mol.\%\,Y        &  \textit{SICCAS}   &   3\,$\times$\,3\,$\times$\,10  & \checkmark  & \rule{0pt}{2.5ex}\\
        BaF$_2$:\,10\,mol.\%\,Y       &  \textit{SICCAS}   &   3\,$\times$\,3\,$\times$\,10  &  \checkmark  &  \rule{0pt}{2.5ex} \\
        BaF$_2$          &  \textit{SICCAS}             &   2\,$\times$\,2\,$\times$\,10      &  & \checkmark    \rule{0pt}{5.0ex}\\
        BaF$_2$:\,3\,mol.\%\,Y         &  \textit{SICCAS}   &   2\,$\times$\,2\,$\times$\,10  &  &  \checkmark    \rule{0pt}{2.5ex} \\
        \bottomrule

    \end{tabularx}
    }
    
\end{table}

\begin{table}[] {\small 
\caption{Summary of the material producers, dimensions and physical and scintillation properties of the samples tested in this work. \ce{BaF2} light output and decay times were measured for this study and are discussed in Sec.~\ref{sec:baf2}. 
}
    \label{tab:Table2}
    \centering
    \begin{tabularx}{\textwidth}{@{\extracolsep{\fill}}lllll}
        \toprule
           
        \textbf{Sample}    & \textbf{Density}   & \textbf{Refractive}  & \textbf{Decay time(s)} & \textbf{Light output} \rule{0pt}{3ex}\\
        \textbf{material} &             [g/cm$^3$]          & \textbf{index}       & [ns]                   & [ph/MeV]             \rule[-1.5ex]{0pt}{0ex} \\       
        \midrule
        
        LYSO:Ce \cite{Gundacker-2016a}            & 7.1    & 1.81  & 24 (15\,\%)\enspace 45 (85\,\%)        & 27\,000 \rule{0pt}{2.5ex} \\
        LSO:Ce,Ca \cite{Gundacker-2016a}        & 7.4    & 1.81  & 8 (6\,\%)\enspace 33 (94\,\%)          & 22\,000 \rule{0pt}{2.5ex}\\
        GFAG \cite{Martinazzoli-2021}         & 6.6    & 1.92  & 41 (65\,\%)\enspace 172 (35\,\%)          & 32\,000 \rule{0pt}{5.0ex}\\
        GAGG:Ce:Mg Accelerated \cite{Martinazzoli-2022}    & 6.6    & 1.92  & 1.3 (8\,\%)\enspace 10 (43\,\%)\enspace 42 (49\,\%)         & 17\,000 \rule{0pt}{2.5ex}\\
        EJ232 \cite{EJ232}                    & 1.0    & 1.58  & 1.6 (100\,\%)        & 8\,400  \rule{0pt}{5.0ex}\\
        BGO \cite{Cala-2022}            & 7.1    & 2.15  & 2.0 (1\,\%)\enspace 42 (7\,\%)\enspace 337 (92\,\%)        & 9\,300  \rule{0pt}{5.0ex}\\
        BGSO \cite{Cala-2022}            & 6.9    & 2.1  & 2.0 (1\,\%)\enspace 40 (17\,\%)\enspace 174 (82\,\%)        & 3\,100  \rule{0pt}{2.5ex}\\
        BSO \cite{Lucchini_2020,Cala-2022}            & 6.8    & 2.1   & 2.9 (2\,\%)\enspace 27 (10\,\%)\enspace 107 (88\,\%)      & 1\,800  \rule{0pt}{2.5ex}\\
        PWO \cite{Lecoq-2016}          & 8.3    & 2.16  & 6 (100\,\%)       & 100  \rule{0pt}{2.5ex}\\
        PbF$_2$ \cite{Kratochwil-2021}          & 7.8    & 1.77  & -        & -  \rule{0pt}{2.5ex}\\
        BaF$_2$                            & 4.9    & 1.55  & 0.04 (1\,\%)\enspace 0.7 (5\,\%)\enspace 643 (94\,\%)             & 3\,700  \rule{0pt}{5.0ex}\\
        BaF$_2$:\,1\,mol.\%\,Y          & 4.9    & 1.55  & 0.03 (1\,\%)\enspace 0.7 (30\,\%)\enspace 294 (69\,\%)             & 890 \rule{0pt}{2.5ex}\\
        BaF$_2$:\,3\,mol.\%\,Y         & 4.9    & 1.55  & 0.04 (1\,\%)\enspace 0.7 (29\,\%)\enspace 318 (70\,\%)             & 1\,100  \rule{0pt}{2.5ex}\\
        BaF$_2$:\,5\,mol.\%\,Y          & 4.9    & 1.55  & 0.03 (1\,\%)\enspace 0.7 (32\,\%)\enspace 303 (67\,\%)             & 810 \rule{0pt}{2.5ex}\\
        BaF$_2$:\,10\,mol.\%\,Y          & 4.9    & 1.55  & 0.06 (5\,\%)\enspace 0.7 (52\,\%)\enspace 145 (43\,\%)             & 530   \rule{0pt}{2.5ex} \rule[-1.5ex]{0pt}{0ex}\\
        \bottomrule

    \end{tabularx}
    }
    
\end{table}

\section{Testbeam} \label{sec:TB}

\subsection{Testbeam Experimental Setup} \label{sec:tbexpset}

\begin{figure}
    \centering
    \includegraphics[width=0.8\textwidth]{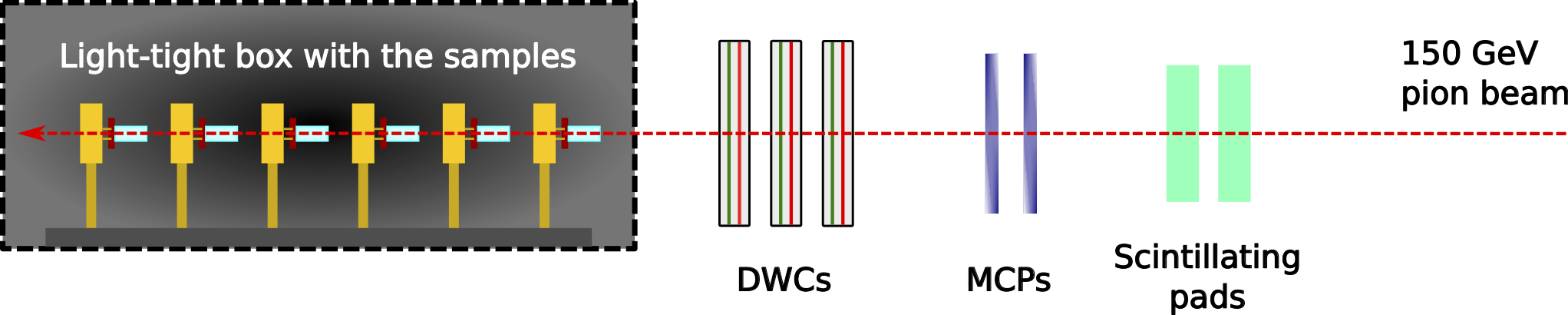}
    \caption{Sketch of the experimental setup. The 150~GeV charged pion beam traverses the samples (light blue) coupled to their SiPMs (brown) and readout boards in a metal case (yellow) placed inside the light-tight box.}
    \label{fig:TB_setup}
\end{figure}
\begin{figure}
    \centering
    \subfigure{
        \includegraphics[width =0.65\textwidth]{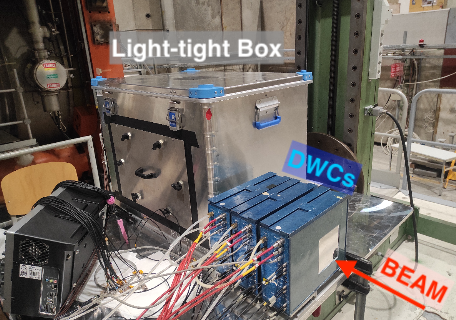}
    }
    \quad \quad
    \subfigure{
        \includegraphics[width =0.25\textwidth]{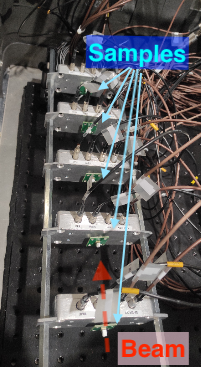}
    }
    \caption{Pictures from the 2023 testbeam. Left: the light-tight box containing the samples with three DWCs (blue) in front. Right: the samples wrapped and connected to the readout boards.}
    \label{fig:TB_photos}
\end{figure}

The measurements were performed at the CERN SPS H2 beamline during two campaigns, in 2022 and 2023. The testbeam set-up is schematically explained in Figure~\ref{fig:TB_setup} and shown in Figure~\ref{fig:TB_photos}.
Two scintillating plastic pads in coincidence provided the hardware trigger, two Microchannel-plates (MCPs) the time reference, three delay wire chambers (DWCs) tracking information. The DWCs were filled with Ar/CO$_2$ gas mixture and read out by a CAEN TDC V1290N.
The MCPs signals were digitized and a timestamp computed for each by constant fraction discrimination (CFD) at 30\%. The average of the 2 timestamps defined the reference time $t_0$ of the event. The resolution of the reference was stable, between 13 and 15~ps standard deviation throughout all the data taking.

The samples were housed in a light-tight box installed on a DESY table moving in the plane orthogonal to the beam direction. Each crystal-SiPM detector was attached to a readout board within a metal shielding case. The 2022 setup allowed the simultaneous measurement of up to six samples, while in 2023 was reduced to five. One measurement slot was occupied at all times by a 2\,$\times$\,2\,$\times$\,3~mm$^3$ LYSO:Ce,Ca sample from FLIR to monitor experimental setup drifts. Throughout the 2023 test beam campaign, the box was maintained at a stable temperature of \SI{16}{\degreeCelsius}.

The SiPM readout board is discussed in \cite{Cates-2018, Gundacker-2019}. The signal was split into two branches, one, fed to an analog operational amplifier and used to compute the deposited energy ("energy" branch), the other to a discrete high-frequency amplifier with bandwidth of approximately 1.5 GHz to compute the timestamp ("time" branch).
All the samples' and the MCPs' pulses were digitized at 5~GS/s with a CAEN V1742 VME module (500~MHz bandwidth) based on the DRS4 chip~\cite{DRS4-2008}.

\subsection{Testbeam Data Analysis}

\begin{figure}
    \centering
    \includegraphics[width = 0.5\textwidth]{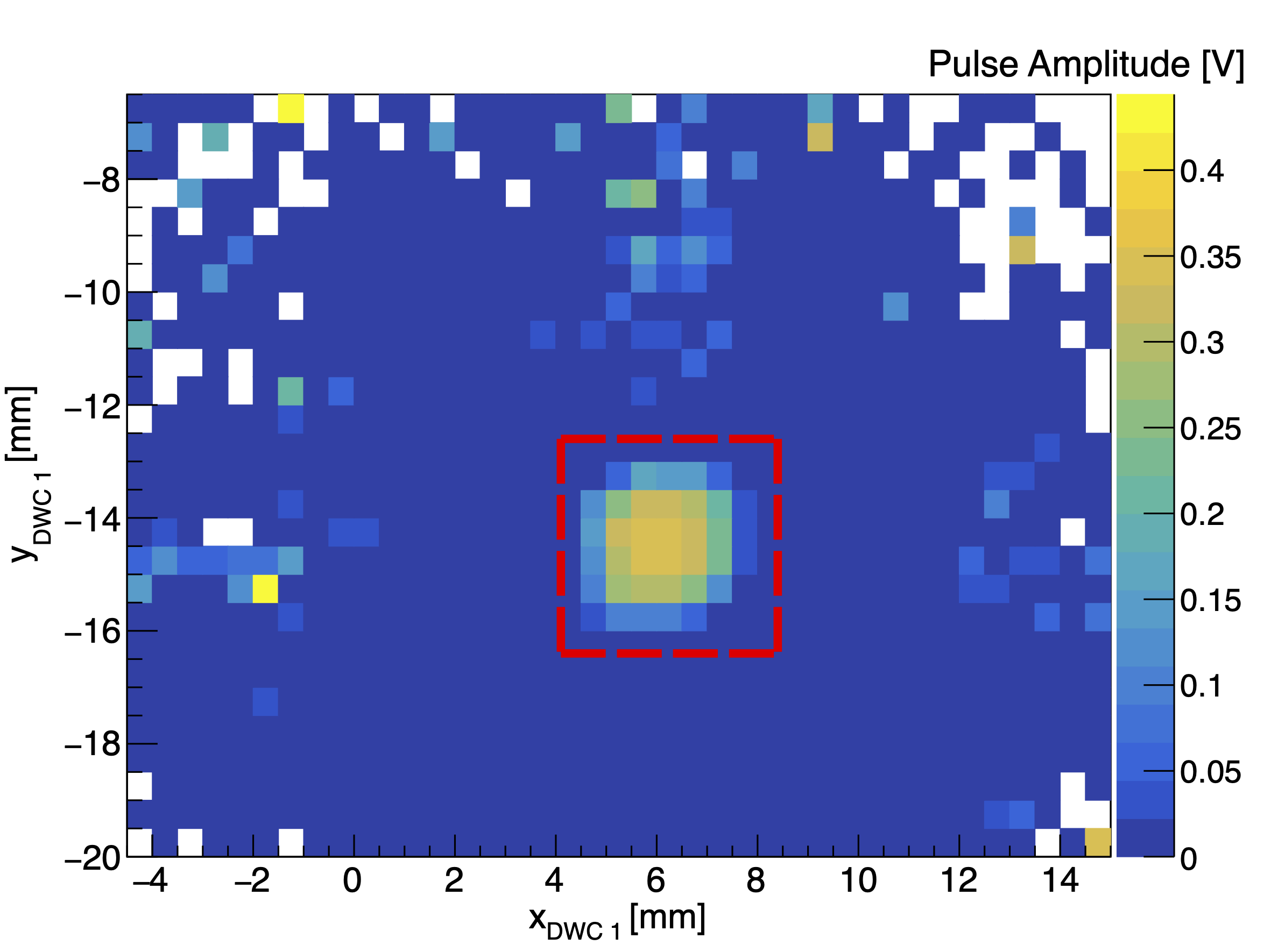}
    \caption{Two-dimensional profile plot illustrating the reconstructed transverse (X, Y) position of the incident beam and the average energy-branch signal amplitude for each (X,Y) bin in the 2\,$\times$\,2\,$\times$\,10~mm$^3$ LYSO:Ce crystal. The red square outlines the crystal. Only events from this region were used for the analysis.}
    \label{fig:DWC_MIPs}
\end{figure}

The signal amplitude of the reference MCPs showed a Landau-like distribution; events below ($<100$~mV) and too high above it ($>300$~mV) were rejected.
Events hitting the samples were identified by plotting a 2D-profile of the reconstructed x and y positions against the energy deposition in the sample (Fig.~\ref{fig:DWC_MIPs}). 
The energy spectra of the samples showed a Landau-like peak with a low-energy tail extending down to 0. Monte Carlo simulations proved it due to particles not traversing the whole length of the sample -- e.g. because of beam divergence and scattering -- which were rejected from the analysis.

The event timestamp of a sample was defined as the time of the leading edge crossing a fixed threshold. The time delay $\Delta t$ is defined as the difference between the event timestamp and the reference time $t_0$ of the MCPs.

\begin{figure}
    \centering
    \subfigure{
    \includegraphics[width=0.49\textwidth]{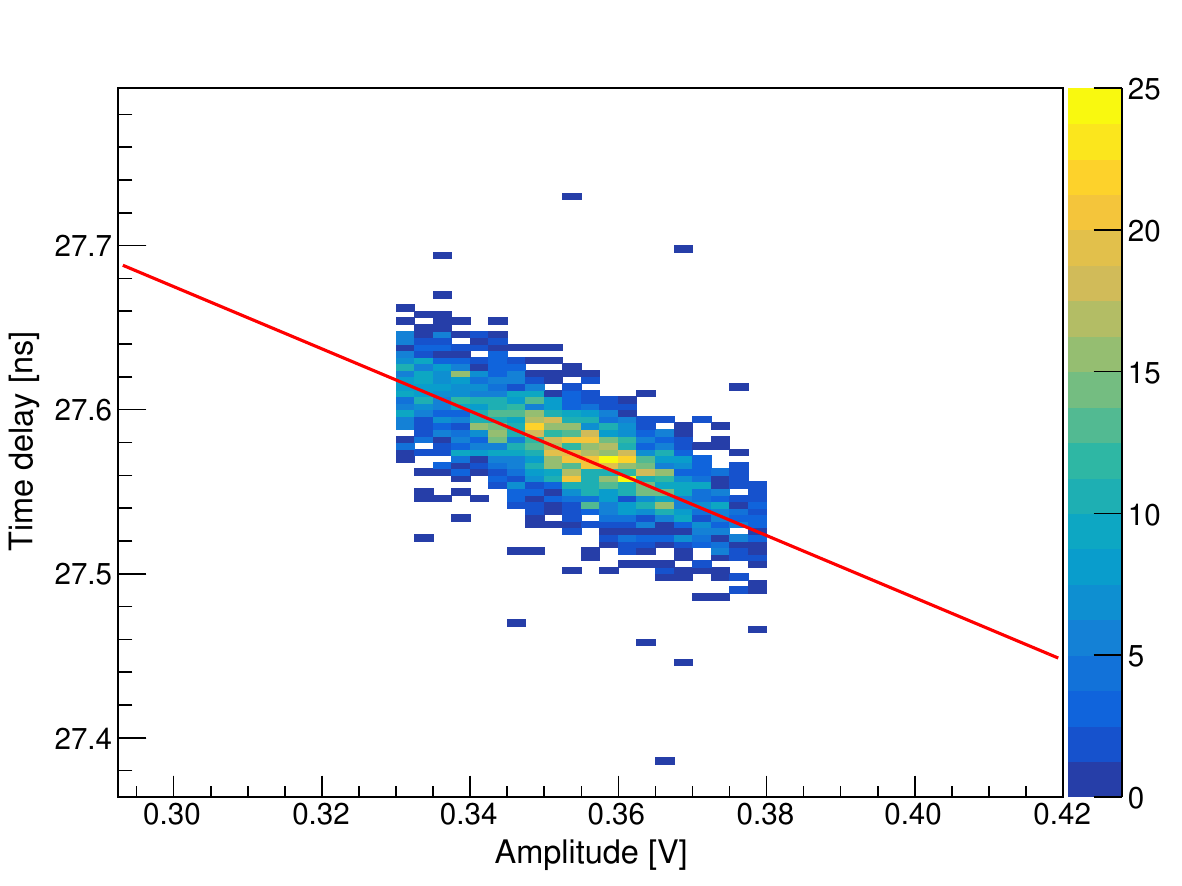}}
    \subfigure{
    \includegraphics[width=0.49\textwidth]{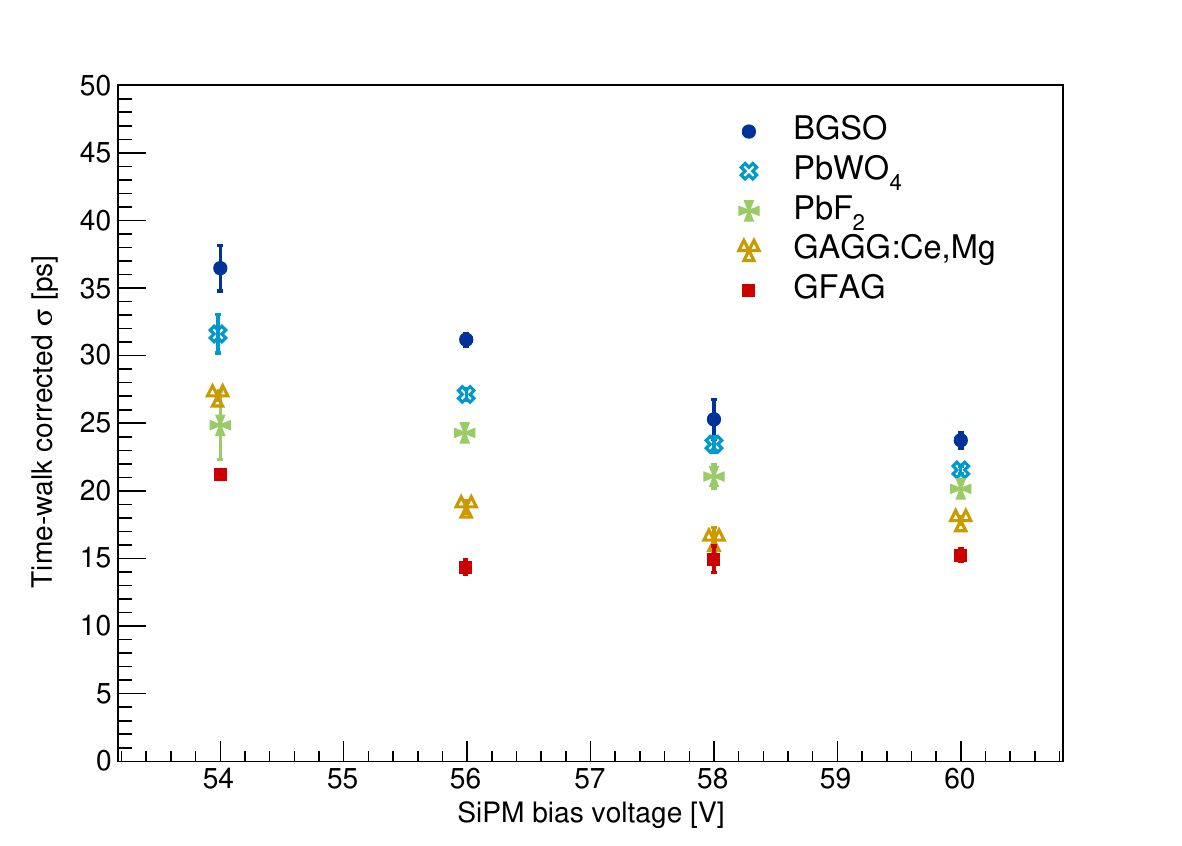}}
    \caption{Left: Two-dimensional histogram of the time delay $\Delta t$ against the energy branch pulse amplitude for the LYSO:Ce sample and 400\,mV threshold on the time branch. The red curve is a linear fit to the data, used to correct for the time-walk.
    Right: Time-walk corrected time resolution of some crystals against the SiPM bias voltage.}
    \label{fig:TWC_MIPs}
    \label{fig:VoltageScan_MIPs}
\end{figure}

\begin{figure}[!h]

\centering
\subfigure{
\includegraphics[width=0.49\textwidth]{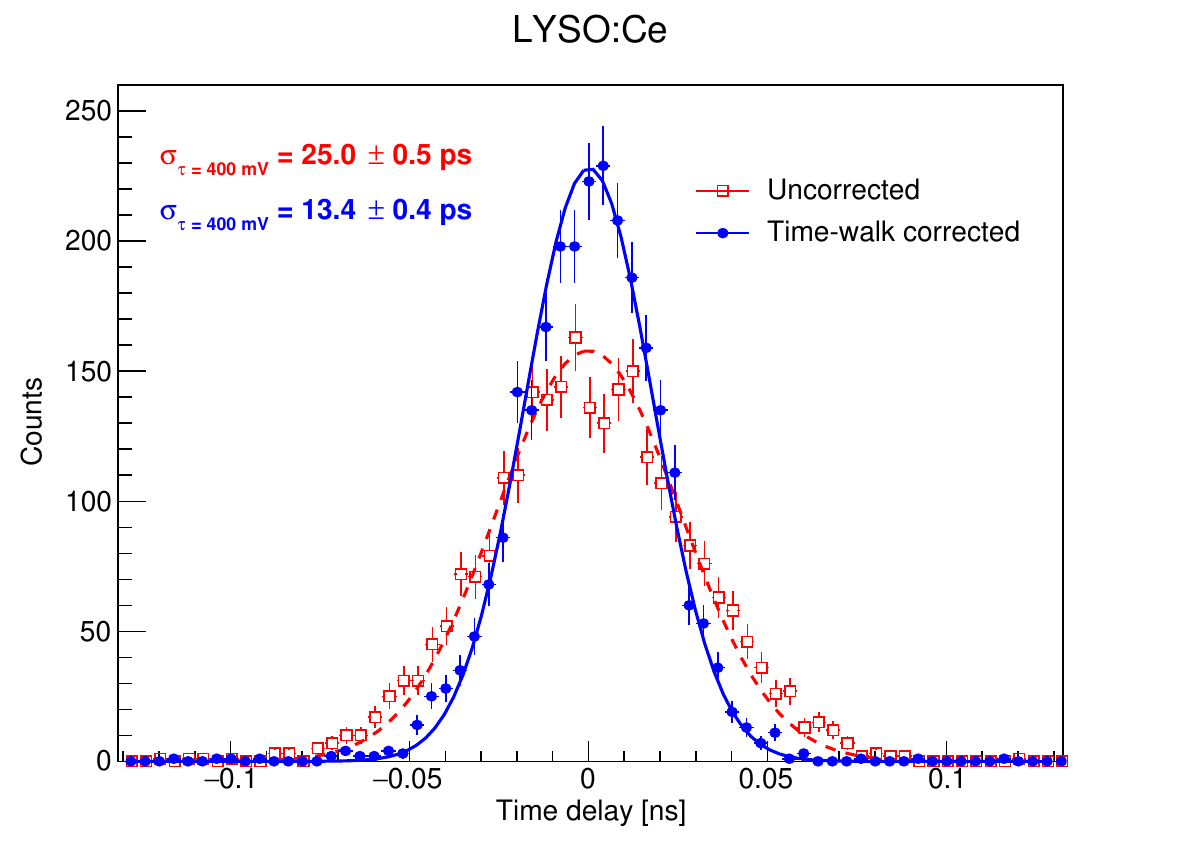}}
\subfigure{
\includegraphics[width=0.49\textwidth]{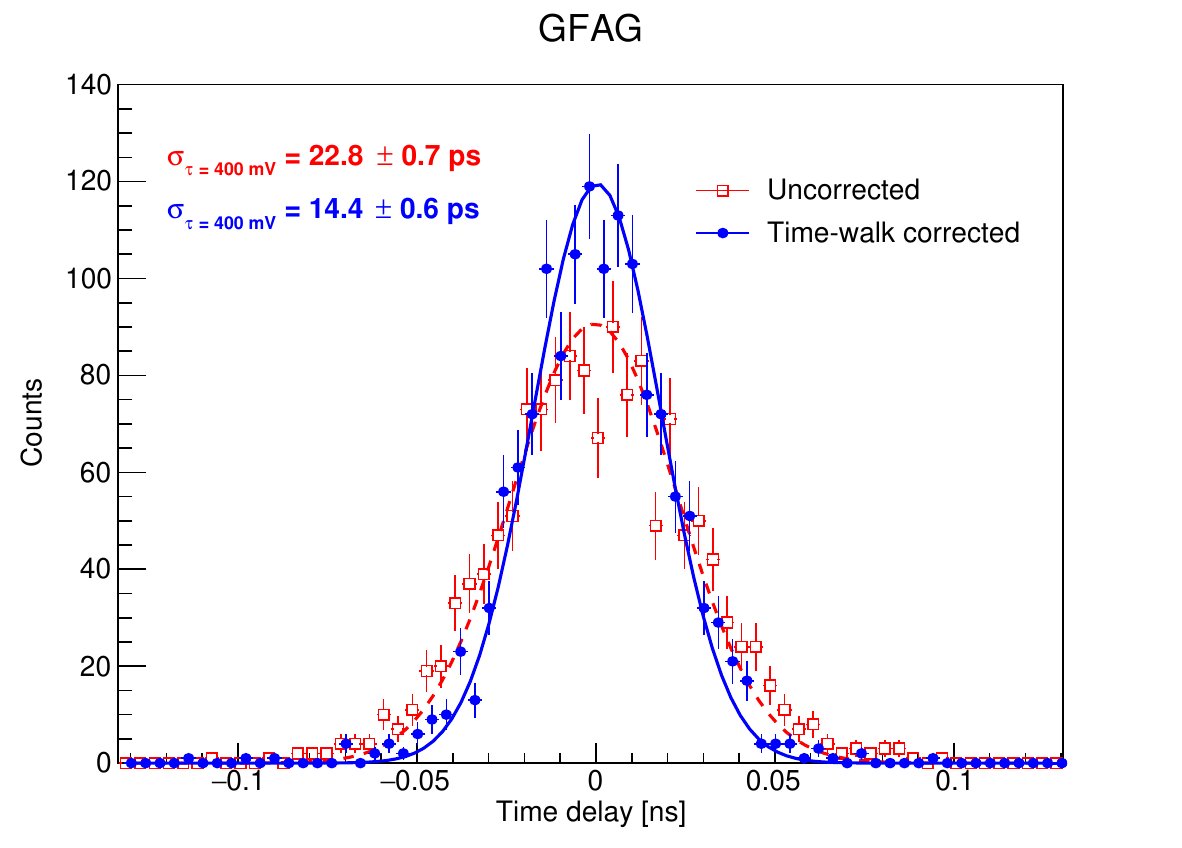}}

\subfigure{
\includegraphics[width=0.49\textwidth]{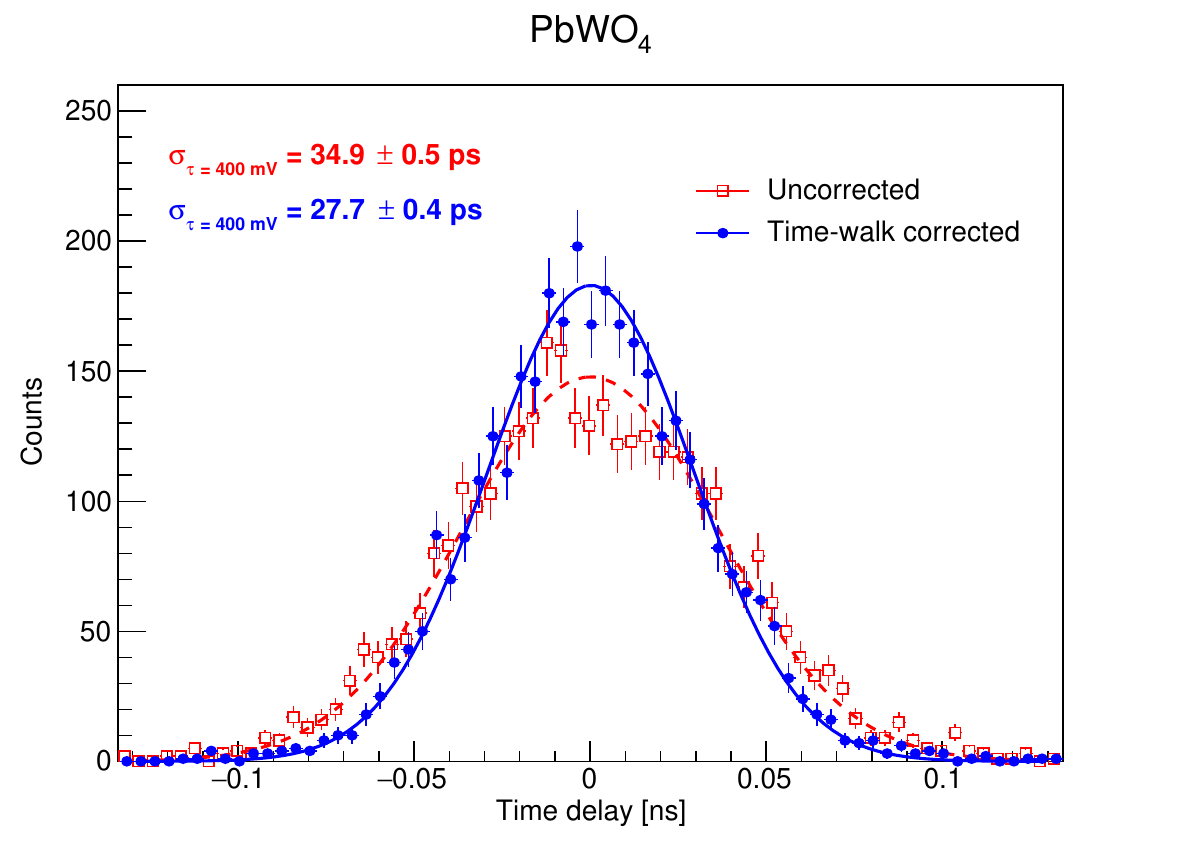}}
\subfigure{
\includegraphics[width=0.49\textwidth]{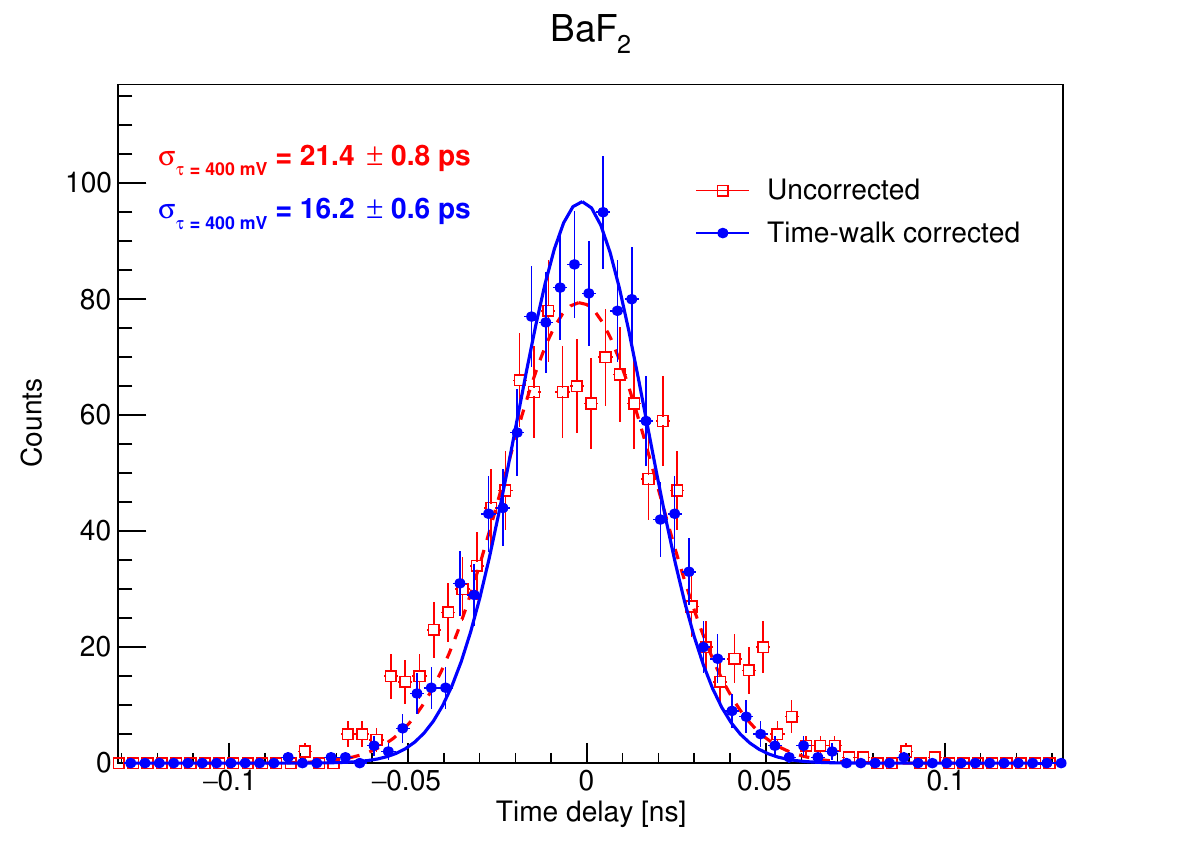}}

\caption{Time delay distributions for some of the samples tested both with and without time-walk correction, and with a Gaussian fit to the data.}
\label{fig:DelayHisto_MIPs}
\end{figure}

The fixed-threshold discrimination technique is affected by time-walk effect due to the fluctuations of the energy deposition and, consequently, of the signal amplitude, visibly correlated to the timestamp -- see Fig.~\ref{fig:TWC_MIPs}, and~\cite{Benaglia-2016, Lucchini-2017}. 
We modelled this relation with a linear function and used the best-fit line to correct the time delay distributions, with a notable enhancement in time performance for many samples (Fig.~\ref{fig:DelayHisto_MIPs}).

\begin{figure}
    \centering
    \subfigure{
    \includegraphics[width=0.49\textwidth]{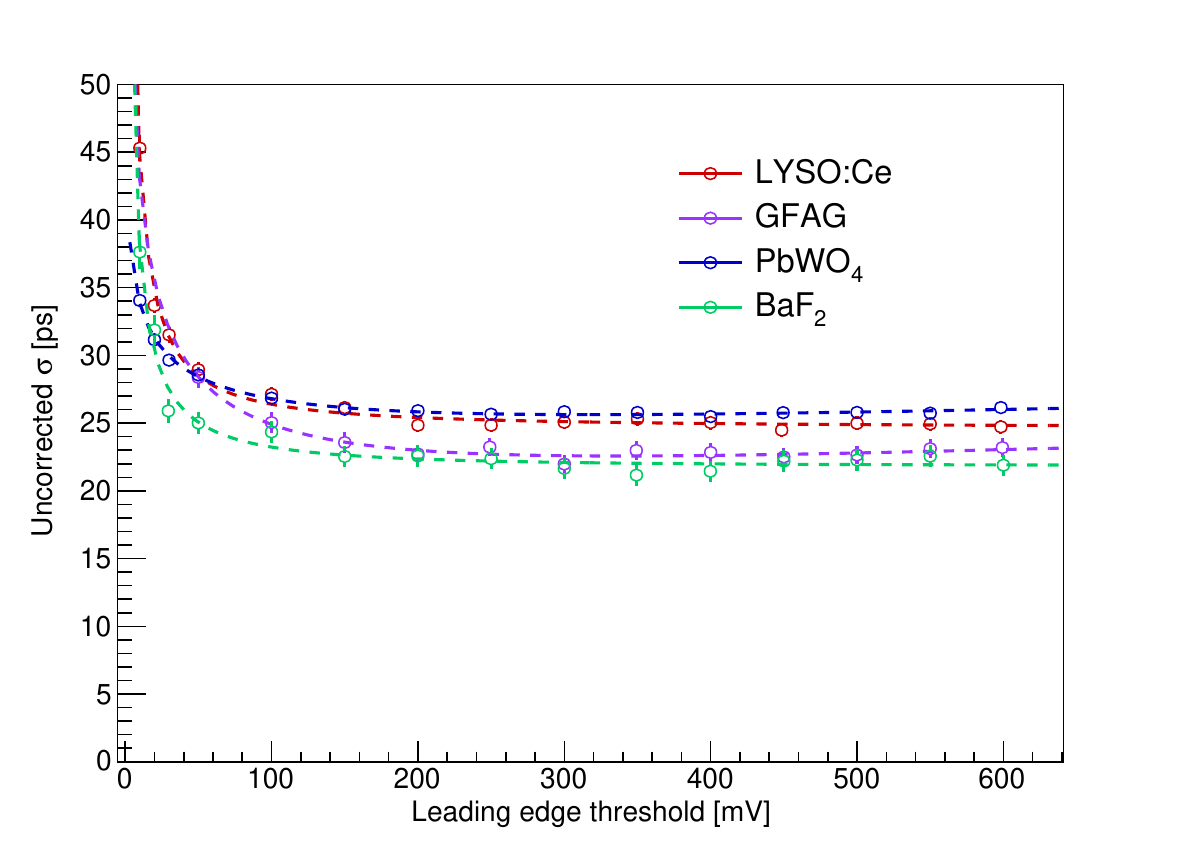}}
    \subfigure{
    \includegraphics[width=0.49\textwidth]{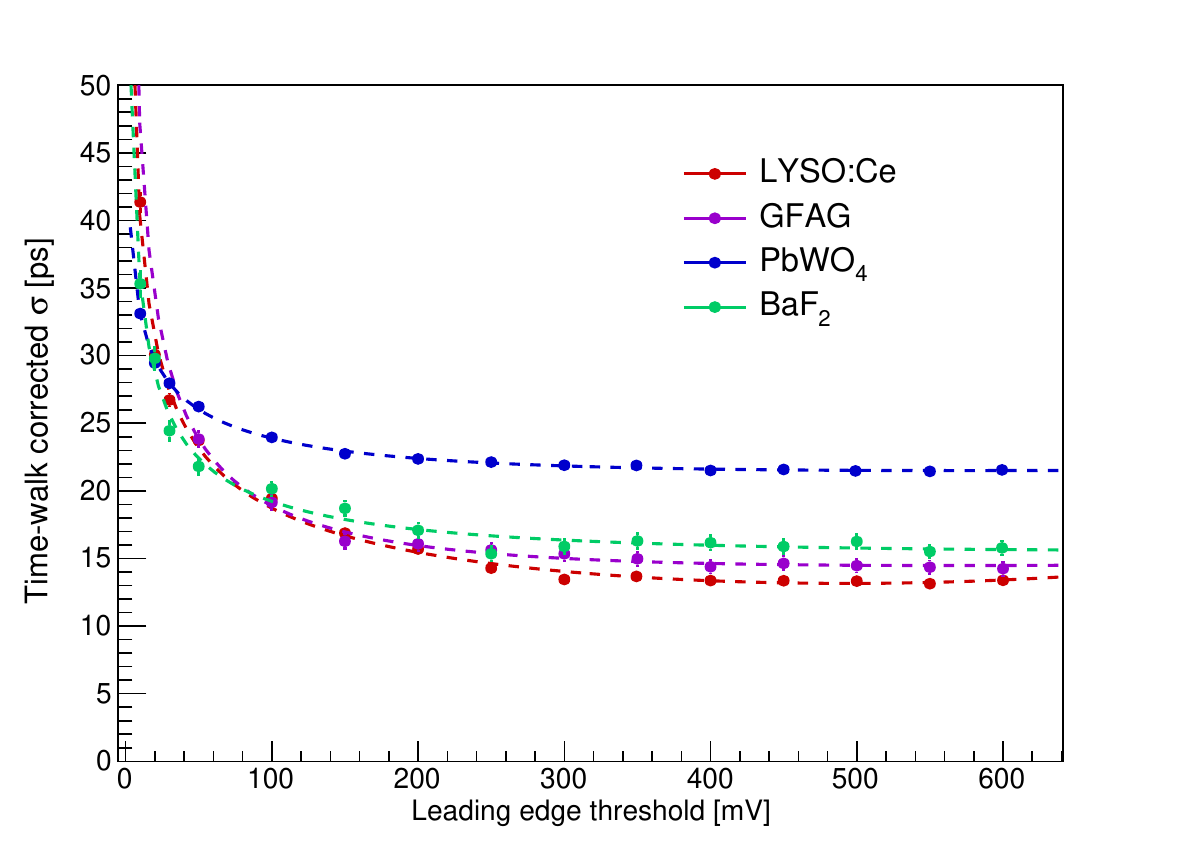}}
    \caption{Leading edge threshold scan of the uncorrected (left) and time-walk corrected (right) time resolution for some of the tested samples.}
    \label{fig:Thr_scan_MIPs}
\end{figure}

The distribution of the time delays $\Delta t$ were well described by a Gaussian for all the samples. The time resolution of the sample was defined as the standard deviation of a Gaussian fit to the distribution with the resolution of the reference MCPs subtracted in quadrature, as in:
\begin{equation*}
    \sigma_\text{t,sample} = \sqrt{\sigma_\text{t, meas}^2 - \sigma_\text{t, MCPs}^2}  
\end{equation*}
For each sample, we selected the threshold that minimised the resolution (Fig.~\ref{fig:Thr_scan_MIPs}).

The SiPM bias was studied with the garnets, BGSO, PWO, and PbF$_2$ coupled to S13360-3050PE SiPMs; Fig.~\ref{fig:VoltageScan_MIPs}, right, shows the time-walk corrected resolution plotted against the SiPM bias voltage.
The timing performance improves with higher SiPM overvoltage, mainly thanks to the increasing PDE. However, at bias voltages above 58~V the dark count rate was visibly intensifying with the integrated pion flux -- noticeably even during a single run. As a compromise, the S13360-3050PE SiPMs were biased at 56~V, while the S13370-6050CN and S13370-3050CN SiPMs for \ce{BaF2} at 58~V.

\section{Testbeam Results}
\label{sec:TBRes}
\subsection{Time Resolution}

\begin{table}[] {\small 
    \centering
    \caption{Summary of the time resolutions measured -- uncorrected and time-walk corrected -- for the samples tested. The average energy deposited by a 150~GeV charged pion is computed with the Monte Carlo simulations (Sec.~\ref{sec:MC}). If the sample was tested both in 2022 and 2023, the values are reported on two separate lines.}    
    \label{tab:TB_res_MIPs}
    \begin{tabularx}{\textwidth}{@{\extracolsep{\fill}}lccccc}
        \toprule        
        \textbf{Sample}  & \textbf{Testbeam} & \textbf{Size} [mm$^3$] &  \textbf{E$_{\textrm{ dep}}$} [MeV]  & \textbf{$\sigma_{\textrm{ uncorr}}$} [ps] & \textbf{$\sigma_{\textrm{ TW corr}}$} [ps]      \rule{0pt}{3ex} \rule[-1.5ex]{0pt} {0ex}\\
        \midrule 

        LYSO:Ce                    & 2022 &   2\,$\times$\,2\,$\times$\,10   & 9.9  &  24.7\,$\pm$\,0.5  &  13.1\,$\pm$\,0.4 \rule{0pt}{2.5ex}\\
        LSO:Ce,Ca                  & 2022 &   2\,$\times$\,2\,$\times$\,10   & 10.3 &  19.4\,$\pm$\,0.4  &  12.1\,$\pm$\,0.4 \rule{0pt}{2.5ex}\\

        GFAG                       & 2022 &   2\,$\times$\,2\,$\times$\,10   & 9.6  &  22.6\,$\pm$\,0.7  &  14.3\,$\pm$\,0.6 \rule{0pt}{5.0ex}\\
                                   & 2023 &   2\,$\times$\,2\,$\times$\,10   & 9.6  &  18.1\,$\pm$\,0.8  &  14.1\,$\pm$\,0.8 \rule{0pt}{2.5ex}\\
        GAGG:Ce:Mg Accelerated 1$^{st}$  & 2022 &   2\,$\times$\,2\,$\times$\,10   & 9.6  &  20.7\,$\pm$\,0.5  &  18.8\,$\pm$\,0.5 \rule{0pt}{2.5ex}\\
        GAGG:Ce:Mg Accelerated 2$^{nd}$  & 2023 &   2\,$\times$\,2\,$\times$\,10   & 9.6  &  19.3\,$\pm$\,1.1  &  13.3\,$\pm$\,0.9 \rule{0pt}{2.5ex}\\

        EJ232                      & 2022 &   3\,$\times$\,3\,$\times$\,3    & 0.5  &  19.8\,$\pm$\,0.2  &  17.2\,$\pm$\,0.2 \rule{0pt}{5.0ex}\\

        BGO                        & 2022 &   2\,$\times$\,2\,$\times$\,10   & 9.8  &  41.4\,$\pm$\,2.1  &  36.4\,$\pm$\,1.5 \rule{0pt}{5.0ex}\\
                                   & 2023 &   2\,$\times$\,2\,$\times$\,10   & 9.8  &  43.5\,$\pm$\,2.1  &  37.9\,$\pm$\,0.9 \rule{0pt}{2.5ex}\\
        BGSO                       & 2022 &   2\,$\times$\,2\,$\times$\,10   & 9.5  &  34.3\,$\pm$\,0.6  &  31.1\,$\pm$\,0.5 \rule{0pt}{2.5ex}\\
                                   & 2023 &   2\,$\times$\,2\,$\times$\,10   & 9.4  &  38.4\,$\pm$\,2.2  &  32.9\,$\pm$\,1.7 \rule{0pt}{2.5ex}\\
        BSO                        & 2023 &   2\,$\times$\,2\,$\times$\,10   & 9.4  &  38.4\,$\pm$\,2.2  &  35.7\,$\pm$\,1.3 \rule{0pt}{2.5ex}\\

        PWO                   & 2022 &   2\,$\times$\,2\,$\times$\,10   & 11.1 &  31.4\,$\pm$\,0.5  &  27.0\,$\pm$\,0.4 \rule{0pt}{5.0ex}\\
        PbF$_2$                    & 2022 &   2\,$\times$\,2\,$\times$\,10   & 10.3 &  37.8\,$\pm$\,1.2  &  24.2\,$\pm$\,0.6 \rule{0pt}{2.5ex}\\

        BaF$_2$                    & 2022 &   3\,$\times$\,3\,$\times$\,10   & 6.7  &  21.8\,$\pm$\,0.8  &  15.8\,$\pm$\,0.6 \rule{0pt}{5.0ex}\\
        BaF$_2$:\,1\,mol.\%\,Y     & 2022 &   3\,$\times$\,3\,$\times$\,10   & 6.7  &  20.1\,$\pm$\,0.4  &  16.0\,$\pm$\,0.4 \rule{0pt}{2.5ex}\\
        BaF$_2$:\,3\,mol.\%\,Y     & 2022 &   3\,$\times$\,3\,$\times$\,10   & 6.7  &  20.4\,$\pm$\,0.4  &  17.0\,$\pm$\,0.4 \rule{0pt}{2.5ex}\\
        BaF$_2$:\,5\,mol.\%\,Y     & 2022 &   3\,$\times$\,3\,$\times$\,10   & 6.7  &  20.2\,$\pm$\,0.5  &  17.7\,$\pm$\,0.5 \rule{0pt}{2.5ex}\\
        BaF$_2$:\,10\,mol.\%\,Y    & 2022 &   3\,$\times$\,3\,$\times$\,10   & 6.7  &  20.6\,$\pm$\,0.3  &  15.7\,$\pm$\,0.3 \rule{0pt}{2.5ex}\\
        BaF$_2$                    & 2023 &   2\,$\times$\,2\,$\times$\,10   & 6.7  &  16.2\,$\pm$\,0.4  &  14.3\,$\pm$\,0.6 \rule{0pt}{5.0ex}\\
        BaF$_2$:\,3\,mol.\%\,Y     & 2023 &   2\,$\times$\,2\,$\times$\,10   & 6.7  &  17.3\,$\pm$\,1.1  &  13.4\,$\pm$\,0.9 \rule{0pt}{2.5ex}\rule[-1.5ex]{0pt}{0ex}\\
        
        \bottomrule
        
    \end{tabularx}
    }
    
\end{table}

 Both uncorrected and time-walk corrected values of the time resolution of the samples tested in both testbeam campaigns are reported in Table~\ref{tab:TB_res_MIPs}. The results for the samples measured both in 2022 and 2023, namely GFAG, BGO, and BGSO, are compatible.  

 It is remarkable to observe that many materials exhibited sub-20\,ps values.
LYSO:Ce and LSO:Ce,Ca coupled to \textit{Hamamatsu} SiPMs showed improved time resolutions of 13.1 and 12.1\,ps respectively, compared to previous works \cite{Benaglia-2016, Lucchini-2017}. Such difference is due to the read-out electronics employed: as presented in \cite{Gundacker-2019}, the custom high-frequency amplifiers used in this work demonstrated to achieve better time performance than the NINO electronics used in \cite{Benaglia-2016, Lucchini-2017} due to higher bandwidth and, thus, better signal-to-noise ratio.

The highly-doped GAGG samples were expected to perform compatibly with the GFAG but with a decay time almost an order of magnitude faster~\cite{Martinazzoli-2022}. The first sample fell short of it, at 18.8~ps; its internal cracks and its imperfect surface state could explain the discrepancy. However, the second GAGG sample, tested in 2023 and overall better, demonstrated a time resolution of 13.3~ps$\pm$0.9, consistent with GFAG. 

As for the only plastic scintillators tested (EJ232), a time resolution of 17.2~ps was obtained despite the shorter thickness of the sample and consequently the lower energy deposition.

The resolutions of B(GS)O, PWO, and PbF$_2$ ranged from 24 to 36~ps. For these, scintillation is too slow -- or absent in PbF$_2$ -- and performance is driven by Cherenkov photons~\cite{Gundacker_2020, Kratochwil_2020, Kratochwil-2021}; their scarcity, both in production and detection, accounts for timing being poorer than that of the traditional scintillators discussed so far. Amongst the B(GS)O, the best time performance of about 33~ps is measured for the mixed BGSO sample, in agreement with results at 511~keV (see~\cite{Cala-2022}). PWO and PbF$_2$ exhibited the best time performance. The very high refractive index and wide transparency range of PbF$_2$ create favourable conditions for the production and detection of Cherenkov photons.

\begin{figure}
    \centering
    \includegraphics[width = 0.6\textwidth]{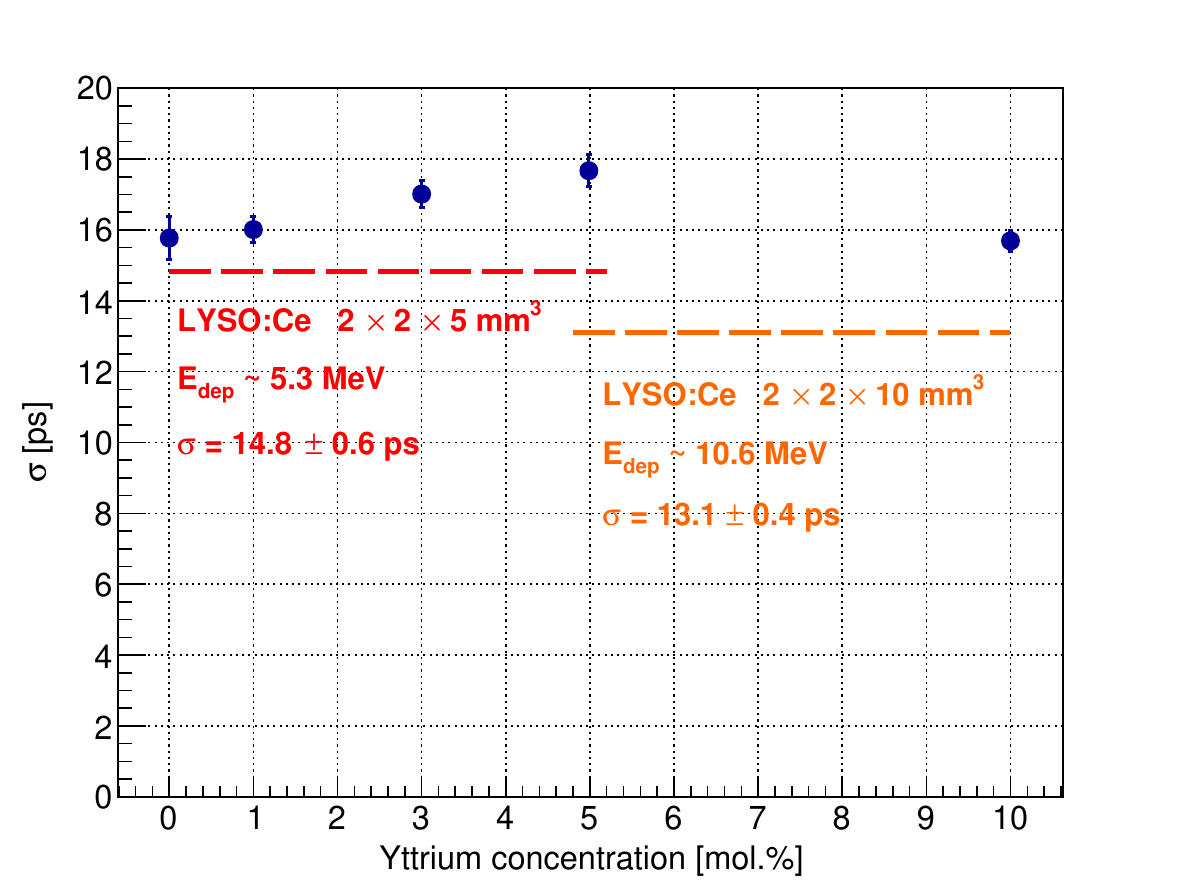}
    \caption{Time resolution against the yttrium doping concentration for the \ce{BaF2} samples. The resolution is mostly unaffected by the dopant and close to the reference LYSO crystals.}
    \label{fig:BaF2paper}
\end{figure}

All the \BaF and \BaFY achieved a time resolution below 20\,ps. The study conducted in 2022 with different amounts of yttrium showed no significant performance variation with the dopant, coherently with laboratory measurements, predictions, and past studies (see Sec.~\ref{sec:baf2} and \cite{Gundacker_2021b}). Small differences among the values can be explained by experimental imperfections such as non-optimal crystal-SiPM coupling or misalignment to the beam. For the 2023 campaign we prepared 2\,$\times$\,2\,$\times$\,10~mm$^3$ BaF$_2$ and \ce{BaF2}:Y samples: both achieved a performance comparable to LYSO of the same dimensions and coupled to a SiPM with the same active area (Fig.~\ref{fig:BaF2paper}). The results is all the more striking considering the low detection efficiency ($\sim$\,15-20\,\%) of their VUV SiPMs compared to the typical values of sensors sensitive at larger wavelengths, which may reach PDEs up to 60\,\% \cite{Gola-2019, Gallina-2019, Gundacker-2019, Gundacker_2021b}.

\subsection{Explorations}
\label{sec:explorations}
\begin{figure}
    \centering
    \includegraphics[width = 0.6\textwidth]{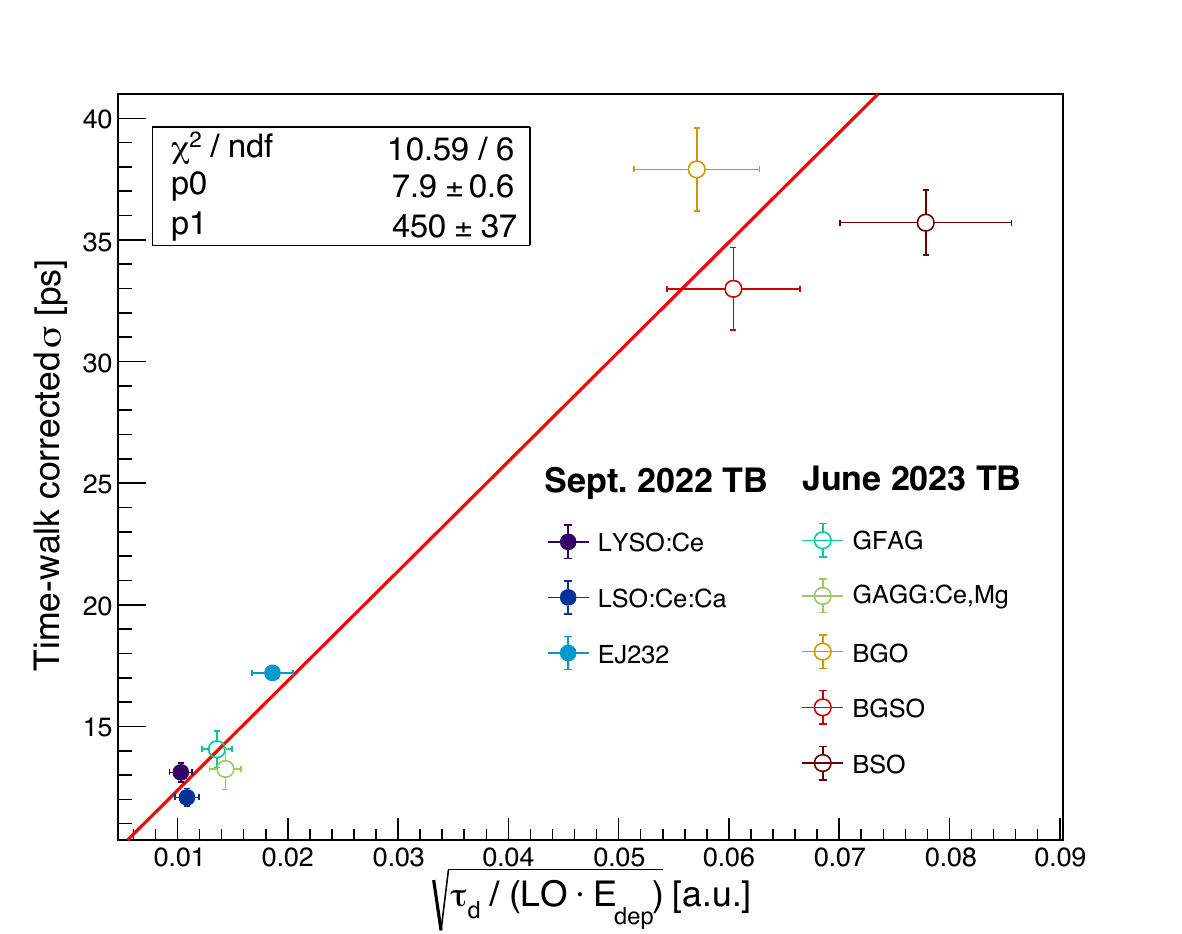}
    \caption{Time-walk corrected time resolutions of the samples against $\sqrt{{\tau_d}/{(LO \cdot E_{dep})}}$, i.e. right hand side of Eq.~\ref{eq:tres}. The points are well described by the model.}
    \label{fig:Correlation}
\end{figure}

The scintillator's contribution to the time resolution of a detector is determined by its scintillation properties: light output, rise time, and decay time.
This dependence has been studied in the past years theoretically -- via order statistics~\cite{Sigfridsson-1967, Seifert_2012} or point processes~\cite{Vinogradov-2018} -- and confirmed experimentally~\cite{Gundacker_2020,Martinazzoli-2021}, showing that the time resolution $\sigma_t$ follows:
\begin{equation}\label{eq:tres}
    \sigma_{t} \propto \sqrt{\frac{\tau_{d,eff}}{LO \cdot E_{dep}}}
\end{equation}
where the effective decay time $\tau_{d,eff}=\left(\sum_i\frac{R_i}{\tau_{d,i}}\right)^{-1}$ is the harmonic mean of the scintillation decay times weighted by the fraction of light emitted via the i-th component, $LO$ the total light output, as function of the energy deposited $E_{dep}$.
Figure\,\ref{fig:Correlation} presents the measured time resolution as a function of the right side of Eq.~\ref{eq:tres}. Data agree with the linear model. The non-zero intercept of the fit line ($7.9\pm0.6$~ps) accounts for contributions beyond the photostatistics. One of such contributions is the random jitter of the digitizer's channels, measured as ($4.0\pm0.3$)~ps; this was obtained by analysing the distribution of the difference in time of arrival of an identical square wave fed into 2 digitizer's channels.

\begin{figure}[!h]
    \centering
    \includegraphics[width = 0.6\textwidth]{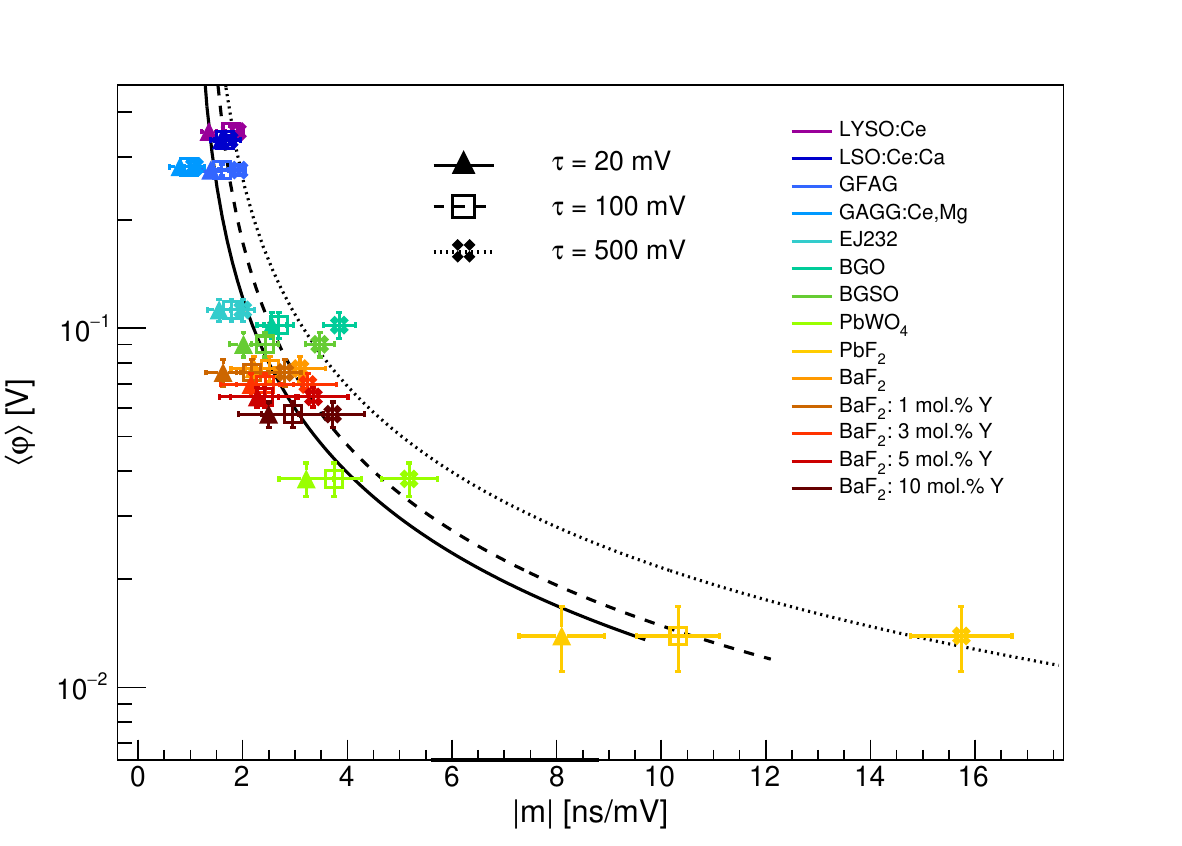}
    \caption{Average energy amplitude $\langle\varphi\rangle$ generated by a charged pion against the unsigned slope $|m|$ of the fit line of the time delay - energy pulse amplitude plot (see Fig.\,\ref{fig:TWC_MIPs} left) at 20, 100 and 500\,mV thresholds on the time signal. The samples with lower light output have steeper slope, i.e. suffer more from time walk.} 
    \label{fig:TWC_MIPs_b}
\end{figure}

Finally, we examined the impact of the time-walk correction on the different samples. From the linear fit of the time delay-energy pulse scatter plot, as the one presented in Fig.~\ref{fig:TWC_MIPs}, the slope values $|m|$ were computed for each sample at 20, 100 and 500~mV leading edge threshold. Their absolute values were then plotted against the average energy pulse amplitude $\langle\varphi\rangle$ (see Fig.\,\ref{fig:TWC_MIPs_b}). As described in \cite{Sigfridsson-1967}, the fitting function of the data points can be approximated as:

\begin{equation}
    m(\langle\varphi\rangle) \simeq p_0 / ( \langle\varphi\rangle + p_1) 
\end{equation}

A good match can be observed between the modelling lines and the data points for the different thresholds.
The effects of the time-walk correction are stronger -- i.e. larger improvements in time resolution -- for the crystals with low light output, e.g. the Cherenkov radiators. This can be explained since the Poisson fluctuations in photons detected relative to the average are larger for lower outputs.

\section{Monte Carlo Simulations} \label{sec:MC}
\subsection{Monte Carlo Simulation Framework}

The testbeam measurements were reproduced with Monte Carlo simulations, including the radiation-matter interaction, the light production and transport, and the formation of the digitised signal. The interaction of a 150~GeV $\pi^+$ beam with the samples coupled to their SiPMs was simulated with the GEANT4 toolkit using the Physics list QGSP\_BERT~\cite{GEANT4_2003, GEANT4_2006, GEANT4_2016}. The material properties were simulated according to Table~\ref{tab:Table2} and the references therein. The Cherenkov and scintillation photons produced were propagated using the GEANT4 ray-tracing library; in particular, the UNIFIED model was employed for the samples' surfaces~\cite{Levin-1996}. Photons were stopped and recorded upon refraction in the SiPM volume. The information of such photons was employed to produce a digitised waveform. A fraction was rejected according to the quantum efficiency and a random delay was added to their time of arrival according to the single photon time resolution (SPTR) of the SiPMs~\cite{Gundacker_2020}. A single photoelectron response pulse was produced for each photons. The event waveform was the resulting sum of all the single photoelectron reponses, sampled at 5~GS/s.

\subsection{Monte Carlo Simulation and Testbeam Results Comparison} \label{sec:Sim_Outputs_MIPs}

\begin{table}\small
    \centering
    \caption{Average energy deposited by the pion beam and time resolution achieved for the crystal materials simulated.}  
    \label{tab:Energy_sigma_sim_MIPs}
    \renewcommand{\arraystretch}{1.2}
    \begin{tabular}{lcc}
        \toprule

    \textbf{Crystal material}  & \textbf{E$_\text{dep}$} & $\sigma$  \\
        \midrule
        {LYSO:Ce}            & 9.9\,$\pm$\,0.1 MeV  & 12.2\,$\pm$\,0.3 ps \\
        {LSO:Ce,Ca}\,\,\,    & 10.3\,$\pm$\,0.1 MeV & 11.8\,$\pm$\,0.3 ps   \\
        {GFAG}               & 9.6\,$\pm$\,0.1 MeV  & 14.7\,$\pm$\,0.4 ps  \\
        {BGO}                & 9.8\,$\pm$\,0.1 MeV  & 36.6\,$\pm$\,0.9 ps\\
        {PWO}           & 11.1\,$\pm$\,0.2 MeV & 29.4\,$\pm$\,0.8 ps \\
        {\BaF}               & 6.7\,$\pm$\,0.1 MeV  & 12.5\,$\pm$\,0.3 ps   \\
        \bottomrule

    \end{tabular}
\end{table}

\begin{figure}
    \centering
    \subfigure{
        \includegraphics[width =0.47\textwidth]{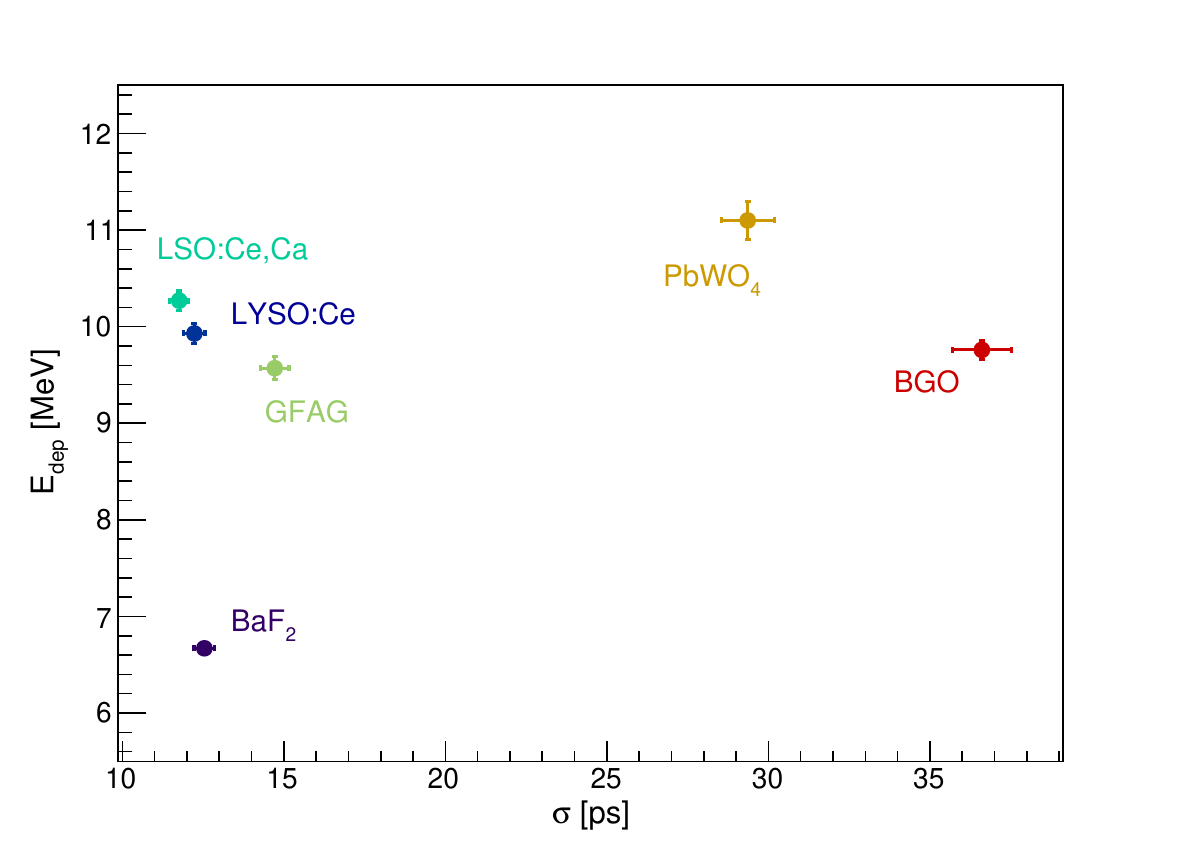}
    }
    \subfigure{
        \includegraphics[width =0.47\textwidth]{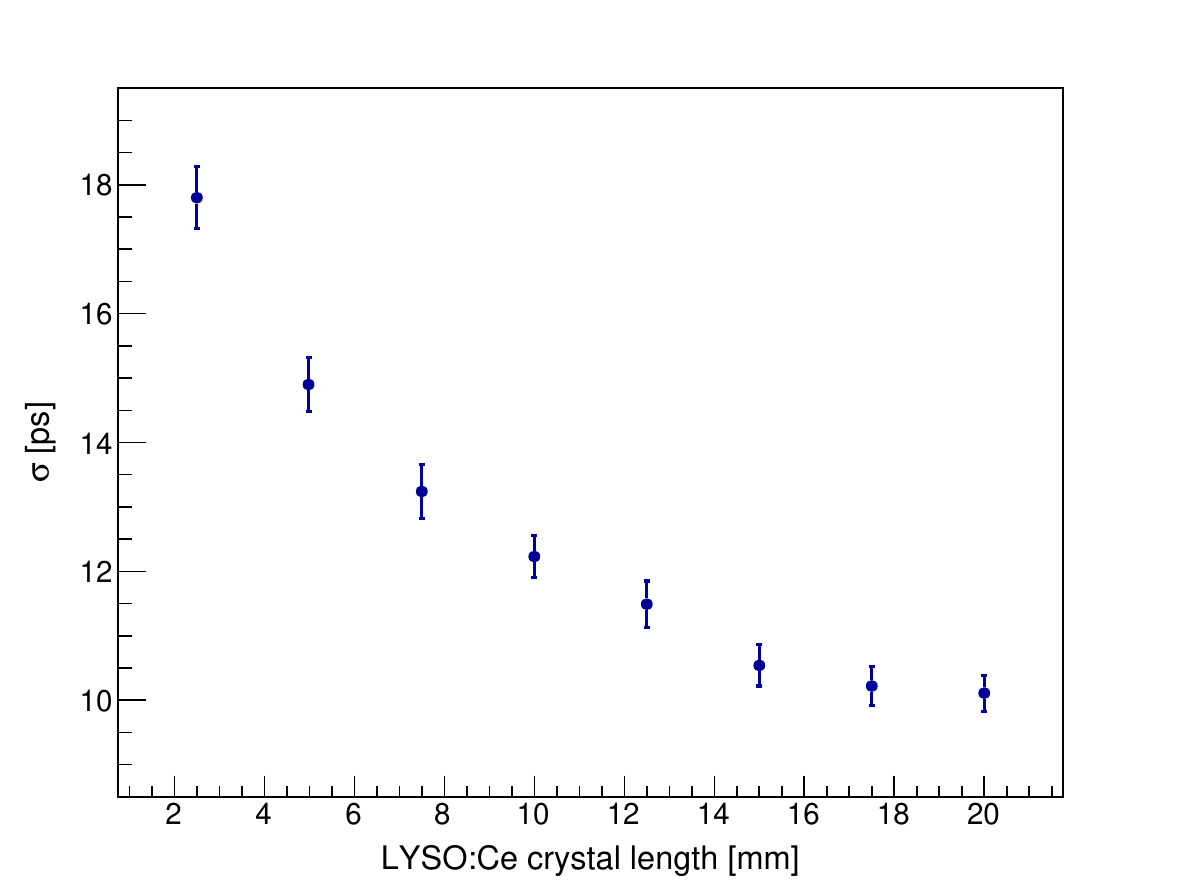}
    }
    \caption{ Left: Monte Carlo simulations, average energy deposited by the pion beam in each simulated crystal against the time performance achieved. Right: simulated time resolution for 2\,$\times$\,2\,$\times$\,$L$~mm$^3$ LYSO:Ce crystals (with $L\in[2,20]$~mm). Longer crystals have better resolution thanks to larger energy deposit and light output; however, the improvement diminishes with the length.}
    \label{fig:sim_MIPs}
\end{figure}

Table~\ref{tab:Energy_sigma_sim_MIPs} shows the average energy deposited and the time resolution for the materials studied in Monte Carlo simulations, also plotted in Fig.~\ref{fig:sim_MIPs} (left); amongst the best performing materials, \BaF offers the least material budget with an average energy deposition of 6.7~MeV.

The right side of Fig.~\ref{fig:sim_MIPs} shows the time resolution achieved by 2\,$\times$\,2~mm$^2$ LYSO:Ce crystals with different lengths in the 2 to 20~mm range. Performance improves for longer crystals thanks to the increasing energy deposits and consequent light production; however, said improvement diminishes with the crystal length because the efficiency of light extraction drops. 
Moreover, half of the scintillation photons are produced in the direction opposite to the SiPM and must undergo at least one reflection at the end opposite to the SiPM before being detected. The additional travel means that these photons will be detected later than the direct ones and most will not contribute to the electronic signal before the threshold is crossed. In other words, these photons do not contribute to the timestamp using a fixed threshold timing technique~\cite{Gundacker_2014}.
This is a major point to consider as longer crystals come at additional cost and material budget in detectors; in this example, doubling the LYSO length from 10 to 20~mm improves performance by less than 20\%.

LSO:Ce,Ca emerged as the top-performing material with time resolution $\sigma_t$ = 11.8\,$\pm$\,0.3~ps, a result consistent with Sec~\ref{sec:Sim_Outputs_MIPs}  and previous experimental studies~\cite{Benaglia-2016, Lucchini-2017}. Generally, the time resolution values for LSO:Ce,Ca, LYSO:Ce, GFAG, and \ce{BaF2} are comparable, while materials such as BGO and PWO, whose timing performance is mainly influenced by Cherenkov light, are considerably worse due to the limited amount of Cherenkov photons detected.

\begin{figure}
    \centering
    \includegraphics[width=0.6\textwidth]{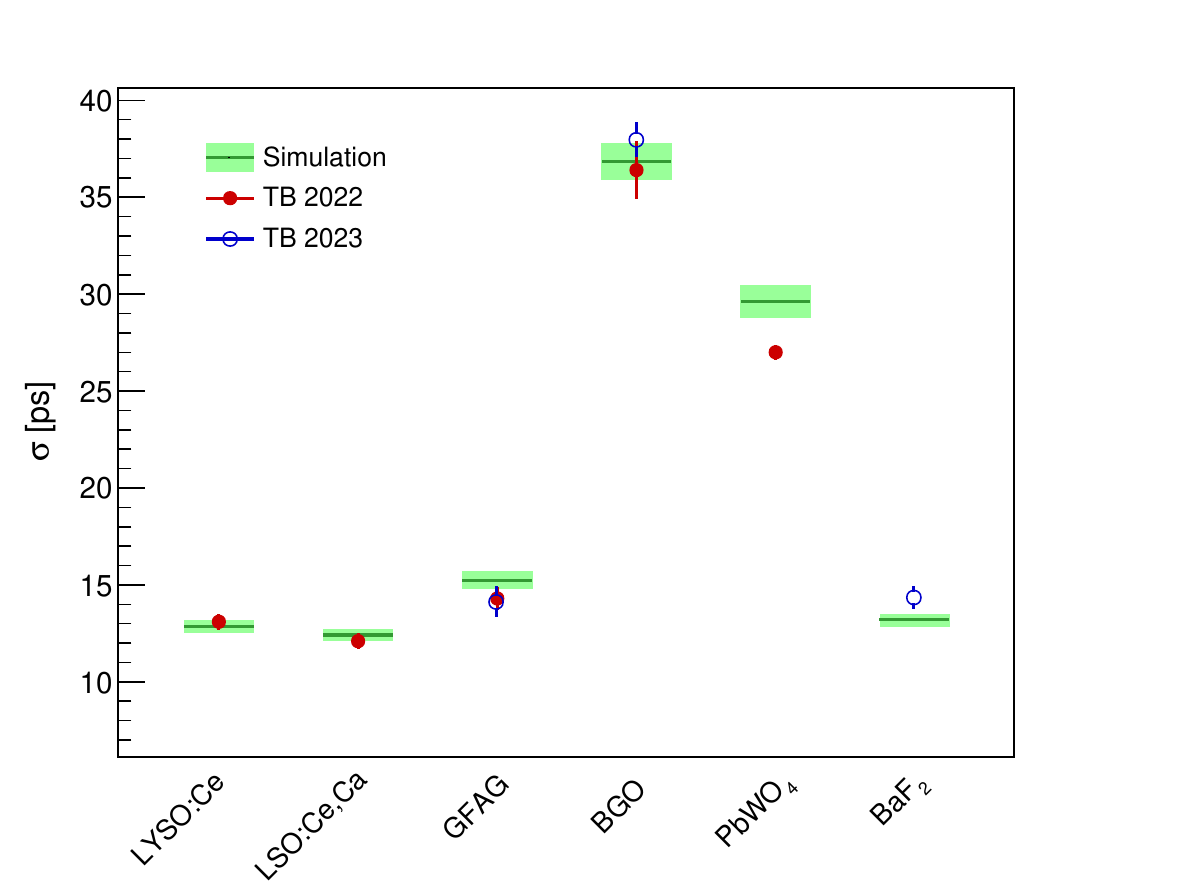}
    \caption{Comparison between the time resolution measured and simulated for some materials under 150~GeV $\pi^+$ irradiation. The simulation values are shown with 1-standard-deviation error bands; a contribution of 4~ps was added in quadrature to account for the electronics contribution.}
    \label{fig:TB_Sim_MIPs}
\end{figure}

Fig.~\ref{fig:TB_Sim_MIPs} shows both simulation and measurement results. A contribution of $\sigma_\text{dig}$ = 4~ps (see Sec.~\ref{sec:explorations}) was added in quadrature to account for the jitter between channels of the digitizer, which was not simulated.
A good overall match is observed between measurements and simulations. The only exception is noted with the PWO sample, where simulations predict a $\sim$3~ps worse value compared to the experimental result. This discrepancy may arise from mismatches between the actual scintillation properties of the sample -- which were impossible to measure with the set-up available due to the low light yield -- and those used for simulations, which were taken from literature.

\section{\BaF and \BaFY Laboratory Characterisation} \label{sec:baf2}
The past sections showed the competitive timing performance of \ce{BaF2}, very close to the best crystals thanks to the sub-nanosecond light emission produced by the cross-luminescence mechanism~\cite{Laval_1983,Lecoq-2017}.
The performance potential is even more evident when considering that the PDE of the SiPMs used is (over) a factor 2 lower than for LSO or garnets~\cite{Gundacker_2020}.
Its radiation length of 2.03~cm is 77\% and 120\% longer than LSO and PWO, respectively,~\cite{PDG} thus making for timing detectors of lower material budget -- i.e. disturbance to detectors behind, for instance an electromagnetic calorimeter~\cite{CMS, CMS2, Lucchini_2020}.

However, \BaF features also a slow scintillation component of $\sim$600~ns linked to self-trapped excitons (STE) recombination. Such component would make it unsuitable in high-rate environments like the High-Luminosity LHC. Recent studies show that it can be quenched with appropriate dopings~\cite{Gundacker_2021}.
In the following sections, we present the laboratory characterisation of the yttrium-doped \BaF crystals tested in the sections above.

\subsection{Crystals and Experimental Setups}

The \BaF samples were produced by the \text{Shanghai Institute of Ceramics}. They presented mirror-polished surfaces, optimal internal state and have 3\,$\times$\,3\,$\times$\,3\,mm$^3$ dimension. Furthermore, to facilitate coincidence time resolution (CTR) measurements with smaller and better performing SiPMs, smaller pixels of sizes 2\,$\times$\,2\,$\times$\,3\,mm$^3$ were cut and mirror-polished at CERN.
The samples were doped using 1, 3, 5 and 10 mol.\% yttrium and compared to undoped \BaF samples.

\subsubsection{Scintillation kinetics under pulsed X-ray excitation}  \label{sec:ScintKin}
Rise and decay time measurements of the tested samples were performed using the Time-Correlated Single Photon Counting (TCSPC) setup illustrated in \cite{pagano_2022}, of which follows a brief account. A \text{PicoQuant} laser model PDL 800-B is employed as an excitation source for a \textit{Hamamatsu} N5084 X-ray tube. The X-ray beam generated had a continuous energy spectrum in the range 0\,-\,40\,keV, with a mean value of $\sim$~10~keV. The beam was focused towards the crystal under test whereas the scintillation light was harvested by a hybrid PMT (model HPM-100-07 from \text{Becker and Hickl}) working in TCSPC mode. The signal of the hybrid PMT provides the stop to a time-to-digital converter (\text{Cronologic} x TDC4), while the time window start is given by an external trigger of the pulsed laser. Starting from the time delays measured, a histogram describing the scintillation kinetics of the crystal is produced.

The scintillation emission-time profile obtained was modelled as a convolution of the impulse response function (IRF) of the setup and a three components bi-exponential function described by the following equation:

\begin{equation}
    f (t | \theta) = \sum_{i=1}^{3} \ R_i \cdot \frac{\textrm{exp}\left( -\frac{t-\theta}{\tau_{d,i}} \right)\ -\ \textrm{exp}\left( -\frac{t - \theta}{\tau_{r}} \right)}{\tau_{d,i} - \tau_{r}} \cdot \Theta\left(t-\theta\right) + bkg
    \label{eq:Rise-decay time}
\end{equation}
being $\tau_{d,i}$ is the $i$-th component of the decay time constant, $\tau_{r}$ is the rise time constant, $R_i$ is the weight of the $i$-th component, $\theta$ is the starting time of the scintillation pulse and $bkg$ is the background.

\subsubsection{Coincidence time resolution setup}

The coincidence time resolution (CTR) of the \BaF samples was measured with the setup explained in detail in \cite{Gundacker-2013, Gundacker-2019}. A $^{22}$Na source emits two correlated 511~keV photons back-to-back which are detected by two crystals coupled to SiPMs, the sample being tested and a reference of known properties. The SiPMs signal is fed into the amplifier board discussed in Sec.~\ref{sec:tbexpset} and the output digitised by a \text{LeCroy} DDA735Zi oscilloscope (3.5~GHz bandwidth, 20~Gs/s sampling rate).

The \BaF samples were measured against a reference  2\,$\times$\,2\,$\times$\,3~mm$^3$ LSO:Ce,Ca\,0.4\% pixel wrapped with several layers of Teflon and coupled to an \text{FBK} NUV-HD SiPM with \text{Cargille} Meltmount optical glue. For this set of measurements, both 2\,$\times$\,2\,$\times$\,3~mm$^3$ and 3\,$\times$\,3\,$\times$\,3~mm$^3$ \BaF and \BaFY pixels were tested coupling them with \text{FBK} and \text{HPK} VUV SiPMs respectively. All \BaF and \BaFY samples were wrapped with several layers of Teflon, dry-coupled to the photodetector and measured in standing position to avoid falling off the SiPM. For the 2\,$\times$\,2\,$\times$\,3~mm$^3$ samples, the face coupled to the sensor was one of the 2~$\times$~2~mm$^2$ surfaces. The \text{FBK} VUV SiPMs were biased at 39\,V ($\sim$\,6~V overvoltage), while the \text{Hamamatsu} ones at 61~V ($\sim$\,9.5~V overvoltage).

A histogram of the mutual time delay between the sample under test and the reference was produced for the photopeak events. The measured CTR ($CTR_{meas}$) was then evaluated as the full width at half maximum (FWHM)\footnote{In accordance with the literature, FWHM is used in this section rather than the Standard Deviation. Note that for a Gaussian distribution FWHM $\approx2.355\sigma$.} of such histogram. Correcting for the reference time resolution ($CTR_{ref}$~=~61\,$\pm$\,3~ps FWHM), the CTR of two identical \BaF samples was calculated as:

\begin{equation}
    CTR_{sample} = \sqrt{2 \cdot CTR_{meas}^2 - CTR_{ref}^2}
\end{equation}

For each sample, a leading edge threshold scan was performed to find the optimal settings on the oscilloscope.

\subsection{Characterisation Results}
\begin{table}[]\small
    \centering
    \caption{Results of Eq.~\ref{eq:Rise-decay time} fitted to the data from the measurements with the TCSPC setup, and CTR measurements. $\tau_{d,i = 1,2,3}$ are the decay time components and $R_{i = 1,2,3}$ their corresponding abundances. An uncertainty of 10\,\%, 6\,\% and 3\,\% can be assumed on the values of $\tau_{d1}$, $\tau_{d2}$ and $\tau_{d3}$ extracted from the distributions, respectively. CTR was measured coupling the  3\,$\times$\,3\,$\times$\,3~mm$^3$ and 2\,$\times$\,2\,$\times$\,3~mm$^3$ samples were coupled to the FBK NUV-HD and Hamamatsu VUV SiPMs, respectively.}    
    \label{tab:BaF2_meas}
    \begin{tabular}{lccccccccc}

        \toprule
        \textbf{Sample} &\textbf{$\tau_{d1}$} [ns] & \textbf{$R_1$} (\%) & \textbf{$\tau_{d2}$} [ns] & \textbf{$R_2$} (\%) & \textbf{$\tau_{d3}$} [ns] & \textbf{$R_3$} (\%) & \multicolumn{2}{c}{CTR (FWHM) @ 511~keV [ps]}   \\
        \midrule
        &&&&&&& 3\,$\times$\,3\,$\times$\,3~mm$^3$ & 2\,$\times$\,2\,$\times$\,3~mm$^3$\\
         
         \midrule
        {\BaF} & 0.038    &	0.5     &	0.707   &	5.2      &	643    &	94.4                                & 71\,$\pm$\,5 & 117\,$\pm$\,5  \\
        {BaF$_2$:~1 mol.\% Y}\,\,\,     & 0.030    &	1.4     &	0.698   &	29.8     &	294    &	68.8    & 81\,$\pm$\,5 & 121\,$\pm$\,5  \\
        {BaF$_2$:~3 mol.\% Y}\,\,\,    & 0.040    &	1.3     &	0.711   &	27.8     &	318    &	70.1        & 71\,$\pm$\,5 & 106\,$\pm$\,5  \\
        {BaF$_2$:~5 mol.\% Y}\,\,\,     & 0.027    &	1.4     &	0.689   &	31.8     &	303    &	66.9    & 82\,$\pm$\,6 & 107\,$\pm$\,5  \\
        {BaF$_2$:~10 mol.\% Y}\,\,\,    & 0.060    &	5.5    &	0.704   &	51.4     &	145    &	43.1    & 76\,$\pm$\,5 & 119\,$\pm$\,5  \\
        \bottomrule
    \end{tabular}

\end{table}

\subsubsection{Scintillation Kinetics}

\begin{figure}[]
\centering
\subfigure[Emission-time distributions - STE emission]{
\includegraphics[width=0.49\textwidth]{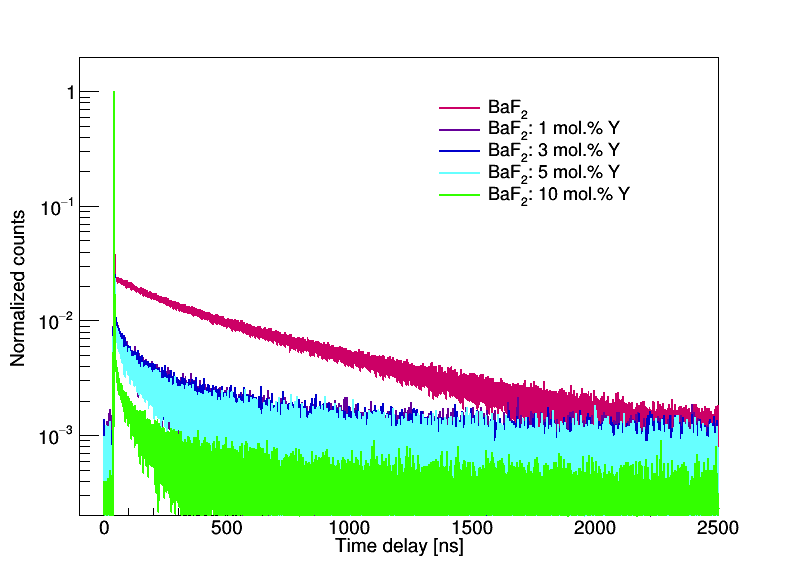}}
\subfigure[Emission-time distributions - Cross-luminescent component]{
\includegraphics[width=0.49\textwidth]{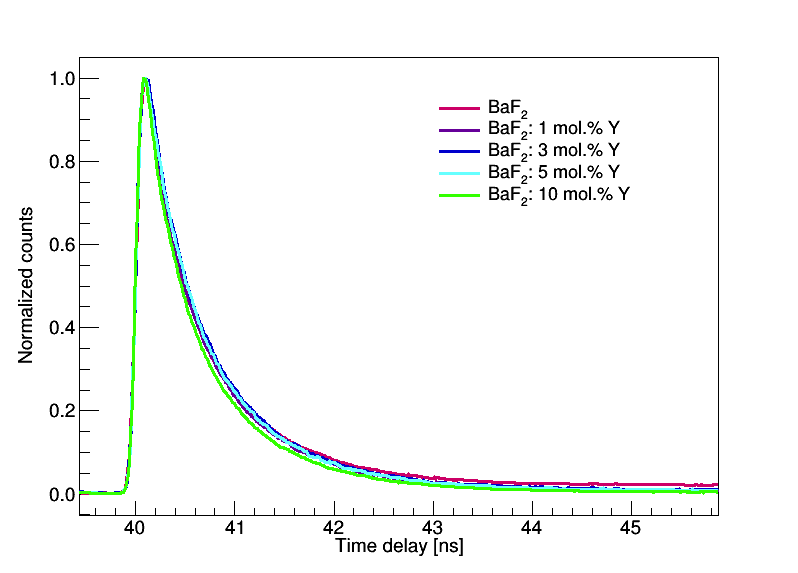}}
\caption{Rescaled scintillation distributions of the \BaF and \BaFY samples under X-ray excitation. Left: plot of the full measurement window. Right: zoom on the rising edge and the cross-luminescence. Increasing the yttrium doping reduces the slow STE emission while keeping constant the fast cross-luminescence.}
\label{fig:BaF2_sup_sk}
\end{figure}

From the measurement performed with the X-ray TCSPC setup, emission-time distributions were obtained for each sample tested (Fig.~\ref{fig:BaF2_sup_sk}). In previous studies \cite{Pots-2020, Gundacker_2021}, it was demonstrated that the cross-luminescent emission of \BaF can be better modelled using two decay time components, one of some tens of picosecond and a second one in the 0.6\,-\,0.8\,ns range. Therefore, all distributions were fitted using a bi-exponential function having three decay time components, two of them to model the fast cross-luminescence emission and a third one for the slow STE component.

The values obtained for the three decay time components and their respective weights are reported in Table\,\ref{tab:BaF2_meas}. The decay times associated with the cross-luminescence do not vary significantly with the increase of yttrium concentration, whereas the STE emission decreases. Therefore, yttrium doping suppresses effectively the slow STE component without affecting the fast cross-luminescent emission. Rise times below the resolution of the setup ($\sim$~20\,ps) were obtained for all the samples.

\subsubsection{Coincidence Time Resolution Measurements}
\begin{figure}[]
    \centering
    \includegraphics[width = 0.6\textwidth]
    {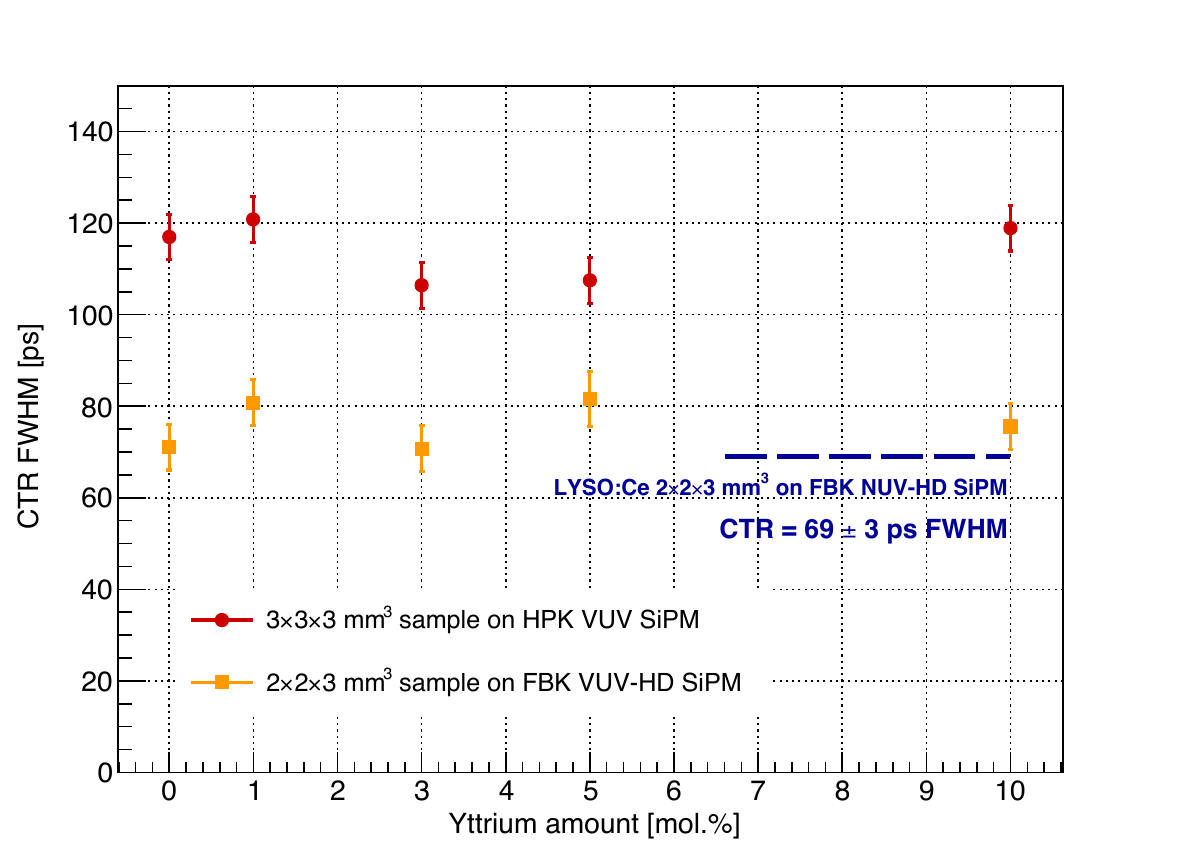}
    \caption{CTR FWHM plotted against the amount of yttrium used as dopant for the two sets of pixels and SiPMs tested in this study.}
    \label{fig:CTR_vs_Y}
\end{figure}

Starting from the events with the 511\,keV $\gamma$-rays full energy deposition in both detectors, it was possible to evaluate the CTR of each sample from the FWHM of the Gaussian function used to model the time delay distribution. To identify the optimum settings, a leading edge threshold scan on the time signal was performed for each sample and the CTR values were extracted from the minimum. 
The CTR FWHM values measured for \BaF and \BaFY samples coupled to {HPK} and {FBK} VUV SiPMs were plotted against the amount of yttrium in the sample in Fig.~\ref{fig:CTR_vs_Y} and reported in Table~\ref{tab:BaF2_meas}.

As in the testbeam measurements, we did not observe any significant variation of the time resolution with the amount of doping.

\clearpage

\section{Conclusions}

This study explored via Monte Carlo simulations and testbeam measurements the timing performance of a variety of materials as light-based MIP timing detectors, including lutetium Orthosilicates, aluminium garnets, Cherenkov radiators, and cross-luminescent crystals.

We obtained the best time resolution with  a 2\,$\times$\,2\,$\times$\,10~mm$^3$ LSO:Ce,Ca coupled to a HPK S13360-3050PE SiPM, 12.1\,$\pm$\,0.4\,ps. 
However, multiple samples achieved a similar time resolution between 13 and 14~ps. Amongst these, we compared: a commercial fast GAGG and a highly-doped GAGG engineered to reduce decay time by almost an order of magnitude; a set of cross-luminescent \BaF doped with yttrium with lower density and less material budget than LYSO. Additionally, materials with timing performance enhanced by Cherenkov photons, like BGSO, PWO, and PbF$_2$, demonstrated noteworthy time performance, with resolution values ranging from 24 to 36~ps.

We measured the scintillation kinetics and the time resolution of the set of \ce{BaF2}:Y, finding significant reduction of the delayed STE scintillation with increasing doping, but without loss of timing performance. 

The results present alternative materials to the traditional LYSO, with competitive performance but additional features: the aluminium garnets offer significantly better radiation tolerance -- tested up to $3\times10^{15}$p/cm$^2$ or 1~MGy (see~\cite{ALENKOV_2019, Lucchini_2016} and \cite{Zhu_2018}) -- while the \BaF lower material budget and potentially ground-breaking performance, particularly with improvements of photon detection in the ultraviolet region where cross-luminescence lies.

\acknowledgments
This work was performed in the framework of the Crystal Clear Collaboration and was supported by the European Union‘s Horizon 2020 Research and Innovation programme under Grant Agreement No 101004761 (AIDAinnova), the Horizon Europe Twinning project TWISMA (grant agreement n°101078960), the CERN EP R\&D Programme on technologies for future experiments, https://ep-rnd.web.cern.ch/ , and by the OP JAC\&MEYS project LASCIMAT – CZ.02.01.01/00/23\_020/0008525.
The authors thank A. Gola from Fondazione Bruno Kessler (FBK) for kindly providing their SiPMs; M. Baschiera, D. Deyrail, Y. Guz, A. Bordelius, and the CERN SPS North Area crew for the technical support.

\newpage
\bibliography{main}

\begin{thebibliography}{48}
\providecommand{\natexlab}[1]{#1}
\providecommand{\url}[1]{\texttt{#1}}
\expandafter\ifx\csname urlstyle\endcsname\relax
  \providecommand{\doi}[1]{doi: #1}\else
  \providecommand{\doi}{doi: \begingroup \urlstyle{rm}\Url}\fi

\bibitem[Group(2020)]{ECFA_2021}
ECFA Detector R\&D Roadmap~Process Group.
\newblock {The 2021 ECFA detector research and development roadmap}.
\newblock Technical report, Geneva, 2020.
\newblock URL \url{https://cds.cern.ch/record/2784893}.

\bibitem[Abada et~al.(2019)]{FCC:2018vvp}
A.~Abada et~al.
\newblock {FCC-hh: The Hadron Collider}: {Future Circular Collider Conceptual Design Report Volume 3}.
\newblock \emph{Eur. Phys. J. ST}, 228\penalty0 (4):\penalty0 755--1107, 2019.
\newblock \doi{10.1140/epjst/e2019-900087-0}.

\bibitem[Collaboration()]{CMS}
The~CMS Collaboration.
\newblock Technical proposal for the phase-ii upgrade of the compact muon solenoid.
\newblock \emph{CERN-LHCC-2015-010/LHCC-P-008}.

\bibitem[CER(2017)]{CERN-LHCC-2017-011}
{The Phase-2 Upgrade of the CMS Barrel Calorimeters}.
\newblock Technical report, CERN, Geneva, 2017.
\newblock URL \url{https://cds.cern.ch/record/2283187}.
\newblock This is the final version, approved by the LHCC.

\bibitem[LHC(2021)]{LHCB_FTDR}
{Framework TDR for the LHCb Upgrade II: Opportunities in flavour physics, and beyond, in the HL-LHC era}.
\newblock Technical report, CERN, Geneva, 2021.
\newblock URL \url{https://cds.cern.ch/record/2776420}.

\bibitem[Aleksa et~al.(2021)Aleksa, Bedeschi, Ferrari, Sefkow, and Tully]{AleksaFCCee}
Martin Aleksa, Franco Bedeschi, Roberto Ferrari, Felix Sefkow, and Christopher~G. Tully.
\newblock Calorimetry at fcc-ee.
\newblock \emph{The European Physical Journal Plus}, 136\penalty0 (10), 2021.
\newblock ISSN 2190-5444.
\newblock \doi{10.1140/epjp/s13360-021-02034-2}.

\bibitem[An et~al.(2023)An, Auffray, Betti, Dall’Omo, Gascon, Golutvin, Guz, Kholodenko, Martinazzoli, {Mazorra De Cos}, Picatoste, Pizzichemi, Roloff, Salomoni, Sanchez, Schopper, Semennikov, Shatalov, Shmanin, Strekalina, and Zhang]{spacal_2023}
L.~An, E.~Auffray, F.~Betti, F.~Dall’Omo, D.~Gascon, A.~Golutvin, Y.~Guz, S.~Kholodenko, L.~Martinazzoli, J.~{Mazorra De Cos}, E.~Picatoste, M.~Pizzichemi, P.~Roloff, M.~Salomoni, D.~Sanchez, A.~Schopper, A.~Semennikov, P.~Shatalov, E.~Shmanin, D.~Strekalina, and Y.~Zhang.
\newblock Performance of a spaghetti calorimeter prototype with tungsten absorber and garnet crystal fibres.
\newblock \emph{Nuclear Instruments and Methods in Physics Research Section A: Accelerators, Spectrometers, Detectors and Associated Equipment}, 1045:\penalty0 167629, 2023.
\newblock ISSN 0168-9002.
\newblock \doi{https://doi.org/10.1016/j.nima.2022.167629}.
\newblock URL \url{https://www.sciencedirect.com/science/article/pii/S0168900222009214}.

\bibitem[CMS(2019)]{CMS_TL}
Collaboration CMS.
\newblock {A MIP Timing Detector for the CMS Phase-2 Upgrade}.
\newblock Technical report, CERN, Geneva, 2019.
\newblock URL \url{https://cds.cern.ch/record/2667167}.

\bibitem[Lecoq et~al.(2020)Lecoq, Morel, Prior, Visvikis, Gundacker, Auffray, Križan, Turtos, Thers, Charbon, Varela, de~La~Taille, Rivetti, Breton, Pratte, Nuyts, Surti, Vandenberghe, Marsden, Parodi, Benlloch, and Benoit]{Lecoq_2020}
Paul Lecoq, Christian Morel, John~O Prior, Dimitris Visvikis, Stefan Gundacker, Etiennette Auffray, Peter Križan, Rosana~Martinez Turtos, Dominique Thers, Edoardo Charbon, Joao Varela, Christophe de~La~Taille, Angelo Rivetti, Dominique Breton, Jean-François Pratte, Johan Nuyts, Suleman Surti, Stefaan Vandenberghe, Paul Marsden, Katia Parodi, Jose~Maria Benlloch, and Mathieu Benoit.
\newblock Roadmap toward the 10 ps time-of-flight pet challenge.
\newblock \emph{Physics in Medicine \& Biology}, 65\penalty0 (21):\penalty0 21RM01, oct 2020.
\newblock \doi{10.1088/1361-6560/ab9500}.
\newblock URL \url{https://dx.doi.org/10.1088/1361-6560/ab9500}.

\bibitem[\textit{et al.}(2016{\natexlab{a}})]{Benaglia-2016}
A.~Benaglia \textit{et al.}
\newblock Detection of high energy muons with sub-20 ps timing resolution using l(y)so crystals and sipm readout.
\newblock \emph{Nucl Instrum Methods Phys Res Sect A Accel Spectrom Detect Assoc Equip}, 830:\penalty0 30--35, 2016{\natexlab{a}}.

\bibitem[\textit{et al.}(2017{\natexlab{a}})]{Lucchini-2017}
M.~T.~Lucchini \textit{et al.}
\newblock Detection of high energy muons with sub-20 ps timing resolution using l(y)so crystals and sipm readout.
\newblock \emph{Nucl Instrum Methods Phys Res Sect A Accel Spectrom Detect Assoc Equip}, 852:\penalty0 1--19, 2017{\natexlab{a}}.

\bibitem[collaboration et~al.(2021)collaboration, Abbott, Abreu, Addesa, Alhusseini, Anderson, Andreev, Apresyan, Arcidiacono, Arenton, Auffray, Bastos, Bauerdick, Bellan, Bellato, Benaglia, Benettoni, Bertoni, Besancon, Bharthuar, Bornheim, Brücken, Butler, Campagnari, Campana, Carlin, Carniti, Cartiglia, Casarsa, Cerri, Checchia, Chen, Chidzik, Chlebana, Cossutti, Costa, Cox, Dafinei, Guio, Debbins, del Re, Dermenev, Marco, Dilsiz, Petrillo, Dissertori, Dogra, Dosselli, Dutta, Caleb, Madrazo, Fernandez, Ferrero, Flowers, Funk, Gallinaro, Ganjour, Gardner, Geurts, Ghezzi, Gninenko, Golf, Gonzalez, Gotti, Gray, Guilloux, Gundacker, Hazen, Hedia, Heering, Heller, Isidori, Isocrate, Jaramillo, Joyce, Kaadze, Karneyeu, Kim, King, Kopp, Korjik, Koseyan, Kozyrev, Kratochwil, Lazarovits, Ledovskoy, Lee, Lee, Li, Li, Li, Liu, Lu, Lucchini, Lustermann, Madrid, Malberti, Mandjavize, Mao, Maravin, Marlow, Marsh, del Arbol, Marzocchi, Mazza, McMahon, Mechinsky, Meridiani, Mestvirishvili, Minafra, Mohammadi, Monti,
  Moon, Mulargia, Murray, Musienko, Nachtman, Nargelas, Narvaez, Neogi, Neu, Niknejad, Obertino, Ogul, Oh, Ojalvo, Onel, Organtini, Orimoto, Ott, Ovtin, Paganoni, Pandolfi, Paramatti, Peck, Perez, Pessina, Pena, Pigazzini, Radchenko, Redaelli, Rigoni, Robutti, Rogan, Rossin, Rovelli, Royon, Sahin, Sands, Santanastasio, Sarica, Schmidt, Schmitz, Sheplock, Silva, Siviero, Soffi, Sola, Sorrentino, Spiropulu, Spitzbart, Leiton, Staiano, Stuart, Suarez, de~Fatis, Tamulaitis, Tang, Tannenwald, Taylor, Tiras, Titov, Tkaczyk, Tlisov, Tlisova, Tornago, Tosi, Tramontano, Trevor, Tully, Ujvari, Varela, Ventura, Vila, Wamorkar, Wang, Wang, Wayne, Wetzel, White, Winn, Wu, Xie, Ye, Yu, Zhang, Zhang, Zhang, Zhang, and Zhu]{Abbott_2021}
The CMS~MTD collaboration, R.~Abbott, A.~Abreu, F.~Addesa, M.~Alhusseini, T.~Anderson, Y.~Andreev, A.~Apresyan, R.~Arcidiacono, M.~Arenton, E.~Auffray, D.~Bastos, L.A.T. Bauerdick, R.~Bellan, M.~Bellato, A.~Benaglia, M.~Benettoni, R.~Bertoni, M.~Besancon, S.~Bharthuar, A.~Bornheim, E.~Brücken, J.N. Butler, C.~Campagnari, M.~Campana, R.~Carlin, P.~Carniti, N.~Cartiglia, M.~Casarsa, O.~Cerri, P.~Checchia, H.~Chen, S.~Chidzik, F.~Chlebana, F.~Cossutti, M.~Costa, B.~Cox, I.~Dafinei, F.~De Guio, P.~Debbins, D.~del Re, A.~Dermenev, E.~Di Marco, K.~Dilsiz, K.F.~Di Petrillo, G.~Dissertori, S.~Dogra, U.~Dosselli, I.~Dutta, F.~Caleb, C.~Fernandez Madrazo, M.~Fernandez, M.~Ferrero, Z.~Flowers, W.~Funk, M.~Gallinaro, S.~Ganjour, M.~Gardner, F.~Geurts, A.~Ghezzi, S.~Gninenko, F.~Golf, J.~Gonzalez, C.~Gotti, L.~Gray, F.~Guilloux, S.~Gundacker, E.~Hazen, S.~Hedia, A.~Heering, R.~Heller, T.~Isidori, R.~Isocrate, R.~Jaramillo, M.~Joyce, K.~Kaadze, A.~Karneyeu, H.~Kim, J.~King, G.~Kopp, M.~Korjik, O.K. Koseyan, A.~Kozyrev,
  N.~Kratochwil, M.~Lazarovits, A.~Ledovskoy, H.~Lee, J.~Lee, A.~Li, S.~Li, W.~Li, T.~Liu, N.~Lu, M.~Lucchini, W.~Lustermann, C.~Madrid, M.~Malberti, I.~Mandjavize, J.~Mao, Y.~Maravin, D.~Marlow, B.~Marsh, P.~Martinez del Arbol, B.~Marzocchi, R.~Mazza, C.~McMahon, V.~Mechinsky, P.~Meridiani, A.~Mestvirishvili, N.~Minafra, A.~Mohammadi, F.~Monti, C.S. Moon, R.~Mulargia, M.~Murray, Y.~Musienko, J.~Nachtman, S.~Nargelas, L.~Narvaez, O.~Neogi, C.~Neu, T.~Niknejad, M.~Obertino, H.~Ogul, G.~Oh, I.~Ojalvo, Y.~Onel, G.~Organtini, T.~Orimoto, J.~Ott, I.~Ovtin, M.~Paganoni, F.~Pandolfi, R.~Paramatti, A.~Peck, C.~Perez, G.~Pessina, C.~Pena, S.~Pigazzini, O.~Radchenko, N.~Redaelli, D.~Rigoni, E.~Robutti, C.~Rogan, R.~Rossin, C.~Rovelli, C.~Royon, M.Ö. Sahin, W.~Sands, F.~Santanastasio, U.~Sarica, I.~Schmidt, R.~Schmitz, J.~Sheplock, J.C. Silva, F.~Siviero, L.~Soffi, V.~Sola, G.~Sorrentino, M.~Spiropulu, D.~Spitzbart, A.G.~Stahl Leiton, A.~Staiano, D.~Stuart, I.~Suarez, T.~Tabarelli de~Fatis, G.~Tamulaitis, Y.~Tang,
  B.~Tannenwald, R.~Taylor, E.~Tiras, M.~Titov, S.~Tkaczyk, D.~Tlisov, I.~Tlisova, M.~Tornago, M.~Tosi, R.~Tramontano, J.~Trevor, C.G. Tully, B.~Ujvari, J.~Varela, S.~Ventura, I.~Vila, T.~Wamorkar, C.~Wang, X.~Wang, M.~Wayne, J.~Wetzel, S.~White, D.~Winn, S.~Wu, S.~Xie, Z.~Ye, G.B. Yu, G.~Zhang, L.~Zhang, Y.~Zhang, Z.~Zhang, and R.~Zhu.
\newblock Test beam characterization of sensor prototypes for the cms barrel mip timing detector.
\newblock \emph{Journal of Instrumentation}, 16\penalty0 (07):\penalty0 P07023, jul 2021.
\newblock \doi{10.1088/1748-0221/16/07/P07023}.
\newblock URL \url{https://dx.doi.org/10.1088/1748-0221/16/07/P07023}.

\bibitem[\textit{et al.}(2019{\natexlab{a}})]{Gundacker-2019}
S.~Gundacker \textit{et al.}
\newblock High-frequency sipm readout advances measured coincidence time resolution limits in tof-pet.
\newblock \emph{Phys Med Biol}, 64\penalty0 (055012), 2019{\natexlab{a}}.

\bibitem[\textit{et al.}(2022{\natexlab{a}})]{Martinazzoli-2022}
L.~Martinazzoli \textit{et al.}
\newblock Compositional engineering of multicomponent garnet scintillators: towards an ultra-accelerated scintillation response.
\newblock \emph{Mater Adv}, 3:\penalty0 6842--6852, 2022{\natexlab{a}}.

\bibitem[\textit{et al.}(1987)]{Klamra-1987}
W.~Klamra \textit{et al.}
\newblock Properties of optical greases for baf$_{2}$ scintillators.
\newblock \emph{Nucl Instrum Methods Phys Res Sect A Accel Spectrom Detect Assoc Equip}, 254:\penalty0 85--87, 1987.

\bibitem[\textit{et al.}(2020)]{Pots-2020}
R.~H.~Pots \textit{et al.}
\newblock Exploiting cross-luminescence in baf$_2$ for ultrafast timing applications using deep-ultraviolet sensitive hpk silicon photomultipliers.
\newblock \emph{Front Phys}, 8:\penalty0 482, 2020.

\bibitem[\textit{et al.}(2016{\natexlab{b}})]{Gundacker-2016a}
S.~Gundacker \textit{et al.}
\newblock Measurement of intrinsic rise times for various l(y)so and luag scintillators with a general study of prompt photons to achieve 10 ps in tof-pet.
\newblock \emph{Physics in Medicine \& Biology}, 61\penalty0 (7):\penalty0 2802, 2016{\natexlab{b}}.

\bibitem[\textit{et al.}(2021{\natexlab{a}})]{Martinazzoli-2021}
L.~Martinazzoli \textit{et al.}
\newblock Scintillation properties and timing performance of state-of-the-art gd$_3$al$_2$ga$_3$o$_{12}$ single crystals.
\newblock \emph{Nucl Instrum Methods Phys Res Sect A Accel Spectrom Detect Assoc Equip}, 1000:\penalty0 165231, 2021{\natexlab{a}}.

\bibitem[EJ2()]{EJ232}
Eljen technology ej232 datasheet.
\newblock URL \url{https://eljentechnology.com/images/products/data_sheets/EJ-232_EJ-232Q}.

\bibitem[\textit{et al.}(2022{\natexlab{b}})]{Cala-2022}
R.~Cala’ \textit{et al.}
\newblock Characterization of mixed bi$_4$(ge$_x$si$_{1 - x}$)$_3$o$_{12}$ for crystal calorimetry at future colliders.
\newblock \emph{Nucl Instrum Methods Phys Res Sect A Accel Spectrom Detect Assoc Equip}, 1032:\penalty0 166527, 2022{\natexlab{b}}.

\bibitem[Lucchini et~al.(2020)Lucchini, Chung, Eno, Lai, Lucchini, Nguyen, and Tully]{Lucchini_2020}
M.T. Lucchini, W.~Chung, S.C. Eno, Y.~Lai, L.~Lucchini, M.~Nguyen, and C.G. Tully.
\newblock New perspectives on segmented crystal calorimeters for future colliders.
\newblock \emph{Journal of Instrumentation}, 15\penalty0 (11):\penalty0 P11005, nov 2020.
\newblock \doi{10.1088/1748-0221/15/11/P11005}.
\newblock URL \url{https://dx.doi.org/10.1088/1748-0221/15/11/P11005}.

\bibitem[\textit{et al.}(2016{\natexlab{c}})]{Lecoq-2016}
P.~Lecoq \textit{et al.}
\newblock \emph{Inorganic Scintillators for Detector Systems}.
\newblock Springer Cham: Springer International Publishing Switzerland 2017, 2016{\natexlab{c}}.

\bibitem[\textit{et al.}(2021{\natexlab{b}})]{Kratochwil-2021}
N.~Kratochwil \textit{et al.}
\newblock A roadmap for sole cherenkov radiators with sipms in tof-pet.
\newblock \emph{Phys Med Biol}, 66:\penalty0 195001, 2021{\natexlab{b}}.

\bibitem[\textit{et al.}(2013{\natexlab{a}})]{Cates-2018}
J.~W.~Cates \textit{et al.}
\newblock Improved single photon time resolution for analog sipms with front end readout that reduces influence of electronic noise.
\newblock \emph{Phys Med Biol}, 63\penalty0 (185022), 2013{\natexlab{a}}.

\bibitem[Ritt(2008)]{DRS4-2008}
Stefan Ritt.
\newblock Design and performance of the 6 ghz waveform digitizing chip drs4.
\newblock In \emph{2008 IEEE Nuclear Science Symposium Conference Record}, pages 1512--1515, 2008.
\newblock \doi{10.1109/NSSMIC.2008.4774700}.

\bibitem[Gundacker et~al.(2020)Gundacker, Turtos, Kratochwil, Pots, Paganoni, Lecoq, and Auffray]{Gundacker_2020}
Stefan Gundacker, Rosana~Martinez Turtos, Nicolaus Kratochwil, Rosalinde~Hendrika Pots, Marco Paganoni, Paul Lecoq, and Etiennette Auffray.
\newblock Experimental time resolution limits of modern sipms and tof-pet detectors exploring different scintillators and cherenkov emission.
\newblock \emph{Physics in Medicine \& Biology}, 65\penalty0 (2):\penalty0 025001, jan 2020.
\newblock \doi{10.1088/1361-6560/ab63b4}.
\newblock URL \url{https://dx.doi.org/10.1088/1361-6560/ab63b4}.

\bibitem[Kratochwil et~al.(2020)Kratochwil, Gundacker, Lecoq, and Auffray]{Kratochwil_2020}
Nicolaus Kratochwil, Stefan Gundacker, Paul Lecoq, and Etiennette Auffray.
\newblock Pushing cherenkov pet with bgo via coincidence time resolution classification and correction.
\newblock \emph{Physics in Medicine \& Biology}, 65\penalty0 (11):\penalty0 115004, jun 2020.
\newblock \doi{10.1088/1361-6560/ab87f9}.
\newblock URL \url{https://dx.doi.org/10.1088/1361-6560/ab87f9}.

\bibitem[Gundacker et~al.(2021{\natexlab{a}})Gundacker, Pots, Nepomnyashchikh, Radzhabov, Shendrik, Omelkov, Kirm, Acerbi, Capasso, Paternoster, Mazzi, Gola, Chen, and Auffray]{Gundacker_2021b}
S~Gundacker, R~H Pots, A~Nepomnyashchikh, E~Radzhabov, R~Shendrik, S~Omelkov, M~Kirm, F~Acerbi, M~Capasso, G~Paternoster, A~Mazzi, A~Gola, J~Chen, and E~Auffray.
\newblock Vacuum ultraviolet silicon photomultipliers applied to baf2 cross-luminescence detection for high-rate ultrafast timing applications.
\newblock \emph{Physics in Medicine \& Biology}, 66\penalty0 (11):\penalty0 114002, jun 2021{\natexlab{a}}.
\newblock \doi{10.1088/1361-6560/abf476}.
\newblock URL \url{https://dx.doi.org/10.1088/1361-6560/abf476}.

\bibitem[\textit{et al.}(2019{\natexlab{b}})]{Gola-2019}
A.~Gola \textit{et al.}
\newblock Nuv-sensitive silicon photomultiplier technologies developed at fondazione bruno kessler.
\newblock \emph{Sensors}, 19:\penalty0 308, 2019{\natexlab{b}}.

\bibitem[\textit{et al.}(2019{\natexlab{c}})]{Gallina-2019}
G.~Gallina \textit{et al.}
\newblock Characterization of the hamamatsu vuv4 mppcs for nexo.
\newblock \emph{Nucl Instrum Methods Phys Res Sect A Accel Spectrom Detect Assoc Equip}, 940:\penalty0 371–379, 2019{\natexlab{c}}.

\bibitem[Sigfridsson(1967)]{Sigfridsson-1967}
B.~Sigfridsson.
\newblock Theoretical analysis of time resolution in scintillation detectors.
\newblock \emph{Nucl Instrum Methods}, 54:\penalty0 13--28, 1967.

\bibitem[Seifert et~al.(2012)Seifert, van Dam, and Schaart]{Seifert_2012}
Stefan Seifert, Herman~T van Dam, and Dennis~R Schaart.
\newblock The lower bound on the timing resolution of scintillation detectors.
\newblock \emph{Physics in Medicine \& Biology}, 57\penalty0 (7):\penalty0 1797, mar 2012.
\newblock \doi{10.1088/0031-9155/57/7/1797}.
\newblock URL \url{https://dx.doi.org/10.1088/0031-9155/57/7/1797}.

\bibitem[Vinogradov(2018)]{Vinogradov-2018}
S.~Vinogradov.
\newblock Approximations of coincidence time resolution models of scintillator detectors with leading edge discrimination.
\newblock \emph{Nucl Instrum Methods Phys Res Sect A}, 912:\penalty0 149--153, 2018.

\bibitem[Agostinelli et~al.(2003)Agostinelli, Allison, Amako, Apostolakis, Araujo, Arce, Asai, Axen, Banerjee, Barrand, Behner, Bellagamba, Boudreau, Broglia, Brunengo, Burkhardt, Chauvie, Chuma, Chytracek, Cooperman, Cosmo, Degtyarenko, Dell'Acqua, Depaola, Dietrich, Enami, Feliciello, Ferguson, Fesefeldt, Folger, Foppiano, Forti, Garelli, Giani, Giannitrapani, Gibin, {Gómez Cadenas}, González, {Gracia Abril}, Greeniaus, Greiner, Grichine, Grossheim, Guatelli, Gumplinger, Hamatsu, Hashimoto, Hasui, Heikkinen, Howard, Ivanchenko, Johnson, Jones, Kallenbach, Kanaya, Kawabata, Kawabata, Kawaguti, Kelner, Kent, Kimura, Kodama, Kokoulin, Kossov, Kurashige, Lamanna, Lampén, Lara, Lefebure, Lei, Liendl, Lockman, Longo, Magni, Maire, Medernach, Minamimoto, {Mora de Freitas}, Morita, Murakami, Nagamatu, Nartallo, Nieminen, Nishimura, Ohtsubo, Okamura, O'Neale, Oohata, Paech, Perl, Pfeiffer, Pia, Ranjard, Rybin, Sadilov, {Di Salvo}, Santin, Sasaki, Savvas, Sawada, Scherer, Sei, Sirotenko, Smith, Starkov, Stoecker,
  Sulkimo, Takahata, Tanaka, Tcherniaev, {Safai Tehrani}, Tropeano, Truscott, Uno, Urban, Urban, Verderi, Walkden, Wander, Weber, Wellisch, Wenaus, Williams, Wright, Yamada, Yoshida, and Zschiesche]{GEANT4_2003}
S.~Agostinelli, J.~Allison, K.~Amako, J.~Apostolakis, H.~Araujo, P.~Arce, M.~Asai, D.~Axen, S.~Banerjee, G.~Barrand, F.~Behner, L.~Bellagamba, J.~Boudreau, L.~Broglia, A.~Brunengo, H.~Burkhardt, S.~Chauvie, J.~Chuma, R.~Chytracek, G.~Cooperman, G.~Cosmo, P.~Degtyarenko, A.~Dell'Acqua, G.~Depaola, D.~Dietrich, R.~Enami, A.~Feliciello, C.~Ferguson, H.~Fesefeldt, G.~Folger, F.~Foppiano, A.~Forti, S.~Garelli, S.~Giani, R.~Giannitrapani, D.~Gibin, J.J. {Gómez Cadenas}, I.~González, G.~{Gracia Abril}, G.~Greeniaus, W.~Greiner, V.~Grichine, A.~Grossheim, S.~Guatelli, P.~Gumplinger, R.~Hamatsu, K.~Hashimoto, H.~Hasui, A.~Heikkinen, A.~Howard, V.~Ivanchenko, A.~Johnson, F.W. Jones, J.~Kallenbach, N.~Kanaya, M.~Kawabata, Y.~Kawabata, M.~Kawaguti, S.~Kelner, P.~Kent, A.~Kimura, T.~Kodama, R.~Kokoulin, M.~Kossov, H.~Kurashige, E.~Lamanna, T.~Lampén, V.~Lara, V.~Lefebure, F.~Lei, M.~Liendl, W.~Lockman, F.~Longo, S.~Magni, M.~Maire, E.~Medernach, K.~Minamimoto, P.~{Mora de Freitas}, Y.~Morita, K.~Murakami, M.~Nagamatu,
  R.~Nartallo, P.~Nieminen, T.~Nishimura, K.~Ohtsubo, M.~Okamura, S.~O'Neale, Y.~Oohata, K.~Paech, J.~Perl, A.~Pfeiffer, M.G. Pia, F.~Ranjard, A.~Rybin, S.~Sadilov, E.~{Di Salvo}, G.~Santin, T.~Sasaki, N.~Savvas, Y.~Sawada, S.~Scherer, S.~Sei, V.~Sirotenko, D.~Smith, N.~Starkov, H.~Stoecker, J.~Sulkimo, M.~Takahata, S.~Tanaka, E.~Tcherniaev, E.~{Safai Tehrani}, M.~Tropeano, P.~Truscott, H.~Uno, L.~Urban, P.~Urban, M.~Verderi, A.~Walkden, W.~Wander, H.~Weber, J.P. Wellisch, T.~Wenaus, D.C. Williams, D.~Wright, T.~Yamada, H.~Yoshida, and D.~Zschiesche.
\newblock Geant4—a simulation toolkit.
\newblock \emph{Nuclear Instruments and Methods in Physics Research Section A: Accelerators, Spectrometers, Detectors and Associated Equipment}, 506\penalty0 (3):\penalty0 250--303, 2003.
\newblock ISSN 0168-9002.
\newblock \doi{https://doi.org/10.1016/S0168-9002(03)01368-8}.
\newblock URL \url{https://www.sciencedirect.com/science/article/pii/S0168900203013688}.

\bibitem[Allison et~al.(2006)Allison, Amako, Apostolakis, Araujo, Arce~Dubois, Asai, Barrand, Capra, Chauvie, Chytracek, Cirrone, Cooperman, Cosmo, Cuttone, Daquino, Donszelmann, Dressel, Folger, Foppiano, Generowicz, Grichine, Guatelli, Gumplinger, Heikkinen, Hrivnacova, Howard, Incerti, Ivanchenko, Johnson, Jones, Koi, Kokoulin, Kossov, Kurashige, Lara, Larsson, Lei, Link, Longo, Maire, Mantero, Mascialino, McLaren, Mendez~Lorenzo, Minamimoto, Murakami, Nieminen, Pandola, Parlati, Peralta, Perl, Pfeiffer, Pia, Ribon, Rodrigues, Russo, Sadilov, Santin, Sasaki, Smith, Starkov, Tanaka, Tcherniaev, Tome, Trindade, Truscott, Urban, Verderi, Walkden, Wellisch, Williams, Wright, and Yoshida]{GEANT4_2006}
J.~Allison, K.~Amako, J.~Apostolakis, H.~Araujo, P.~Arce~Dubois, M.~Asai, G.~Barrand, R.~Capra, S.~Chauvie, R.~Chytracek, G.A.P. Cirrone, G.~Cooperman, G.~Cosmo, G.~Cuttone, G.G. Daquino, M.~Donszelmann, M.~Dressel, G.~Folger, F.~Foppiano, J.~Generowicz, V.~Grichine, S.~Guatelli, P.~Gumplinger, A.~Heikkinen, I.~Hrivnacova, A.~Howard, S.~Incerti, V.~Ivanchenko, T.~Johnson, F.~Jones, T.~Koi, R.~Kokoulin, M.~Kossov, H.~Kurashige, V.~Lara, S.~Larsson, F.~Lei, O.~Link, F.~Longo, M.~Maire, A.~Mantero, B.~Mascialino, I.~McLaren, P.~Mendez~Lorenzo, K.~Minamimoto, K.~Murakami, P.~Nieminen, L.~Pandola, S.~Parlati, L.~Peralta, J.~Perl, A.~Pfeiffer, M.G. Pia, A.~Ribon, P.~Rodrigues, G.~Russo, S.~Sadilov, G.~Santin, T.~Sasaki, D.~Smith, N.~Starkov, S.~Tanaka, E.~Tcherniaev, B.~Tome, A.~Trindade, P.~Truscott, L.~Urban, M.~Verderi, A.~Walkden, J.P. Wellisch, D.C. Williams, D.~Wright, and H.~Yoshida.
\newblock Geant4 developments and applications.
\newblock \emph{IEEE Transactions on Nuclear Science}, 53\penalty0 (1):\penalty0 270--278, 2006.
\newblock \doi{10.1109/TNS.2006.869826}.

\bibitem[Allison et~al.(2016)Allison, Amako, Apostolakis, Arce, Asai, Aso, Bagli, Bagulya, Banerjee, Barrand, Beck, Bogdanov, Brandt, Brown, Burkhardt, Canal, Cano-Ott, Chauvie, Cho, Cirrone, Cooperman, Cortés-Giraldo, Cosmo, Cuttone, Depaola, Desorgher, Dong, Dotti, Elvira, Folger, Francis, Galoyan, Garnier, Gayer, Genser, Grichine, Guatelli, Guèye, Gumplinger, Howard, Hřivnáčová, Hwang, Incerti, Ivanchenko, Ivanchenko, Jones, Jun, Kaitaniemi, Karakatsanis, Karamitros, Kelsey, Kimura, Koi, Kurashige, Lechner, Lee, Longo, Maire, Mancusi, Mantero, Mendoza, Morgan, Murakami, Nikitina, Pandola, Paprocki, Perl, Petrović, Pia, Pokorski, Quesada, Raine, Reis, Ribon, {Ristić Fira}, Romano, Russo, Santin, Sasaki, Sawkey, Shin, Strakovsky, Taborda, Tanaka, Tomé, Toshito, Tran, Truscott, Urban, Uzhinsky, Verbeke, Verderi, Wendt, Wenzel, Wright, Wright, Yamashita, Yarba, and Yoshida]{GEANT4_2016}
J.~Allison, K.~Amako, J.~Apostolakis, P.~Arce, M.~Asai, T.~Aso, E.~Bagli, A.~Bagulya, S.~Banerjee, G.~Barrand, B.R. Beck, A.G. Bogdanov, D.~Brandt, J.M.C. Brown, H.~Burkhardt, Ph. Canal, D.~Cano-Ott, S.~Chauvie, K.~Cho, G.A.P. Cirrone, G.~Cooperman, M.A. Cortés-Giraldo, G.~Cosmo, G.~Cuttone, G.~Depaola, L.~Desorgher, X.~Dong, A.~Dotti, V.D. Elvira, G.~Folger, Z.~Francis, A.~Galoyan, L.~Garnier, M.~Gayer, K.L. Genser, V.M. Grichine, S.~Guatelli, P.~Guèye, P.~Gumplinger, A.S. Howard, I.~Hřivnáčová, S.~Hwang, S.~Incerti, A.~Ivanchenko, V.N. Ivanchenko, F.W. Jones, S.Y. Jun, P.~Kaitaniemi, N.~Karakatsanis, M.~Karamitros, M.~Kelsey, A.~Kimura, T.~Koi, H.~Kurashige, A.~Lechner, S.B. Lee, F.~Longo, M.~Maire, D.~Mancusi, A.~Mantero, E.~Mendoza, B.~Morgan, K.~Murakami, T.~Nikitina, L.~Pandola, P.~Paprocki, J.~Perl, I.~Petrović, M.G. Pia, W.~Pokorski, J.M. Quesada, M.~Raine, M.A. Reis, A.~Ribon, A.~{Ristić Fira}, F.~Romano, G.~Russo, G.~Santin, T.~Sasaki, D.~Sawkey, J.I. Shin, I.I. Strakovsky, A.~Taborda,
  S.~Tanaka, B.~Tomé, T.~Toshito, H.N. Tran, P.R. Truscott, L.~Urban, V.~Uzhinsky, J.M. Verbeke, M.~Verderi, B.L. Wendt, H.~Wenzel, D.H. Wright, D.M. Wright, T.~Yamashita, J.~Yarba, and H.~Yoshida.
\newblock Recent developments in geant4.
\newblock \emph{Nuclear Instruments and Methods in Physics Research Section A: Accelerators, Spectrometers, Detectors and Associated Equipment}, 835:\penalty0 186--225, 2016.
\newblock ISSN 0168-9002.
\newblock \doi{https://doi.org/10.1016/j.nima.2016.06.125}.
\newblock URL \url{https://www.sciencedirect.com/science/article/pii/S0168900216306957}.

\bibitem[Levin and Moisan(1996)]{Levin-1996}
A.~Levin and C.~Moisan.
\newblock A more physical approach to model the surface treatment of scintillation counters and its implementation into detect.
\newblock In \emph{1996 IEEE Nuclear Science Symposium. Conference Record}, volume~2, pages 702--6. IEEE, 1996.

\bibitem[Gundacker et~al.(2014)Gundacker, Knapitsch, Auffray, Jarron, Meyer, and Lecoq]{Gundacker_2014}
S.~Gundacker, A.~Knapitsch, E.~Auffray, P.~Jarron, T.~Meyer, and P.~Lecoq.
\newblock Time resolution deterioration with increasing crystal length in a tof-pet system.
\newblock \emph{Nuclear Instruments and Methods in Physics Research Section A: Accelerators, Spectrometers, Detectors and Associated Equipment}, 737:\penalty0 92--100, 2014.
\newblock ISSN 0168-9002.
\newblock \doi{https://doi.org/10.1016/j.nima.2013.11.025}.
\newblock URL \url{https://www.sciencedirect.com/science/article/pii/S0168900213015647}.

\bibitem[Laval et~al.(1983)Laval, Moszyński, Allemand, Cormoreche, Guinet, Odru, and Vacher]{Laval_1983}
M.~Laval, M.~Moszyński, R.~Allemand, E.~Cormoreche, P.~Guinet, R.~Odru, and J.~Vacher.
\newblock Barium fluoride — inorganic scintillator for subnanosecond timing.
\newblock \emph{Nuclear Instruments and Methods in Physics Research}, 206\penalty0 (1):\penalty0 169--176, 1983.
\newblock ISSN 0167-5087.
\newblock \doi{https://doi.org/10.1016/0167-5087(83)91254-1}.
\newblock URL \url{https://www.sciencedirect.com/science/article/pii/0167508783912541}.

\bibitem[\textit{et al.}(2017{\natexlab{b}})]{Lecoq-2017}
P.~Lecoq \textit{et al.}
\newblock \emph{Inorganic scintillators for detector systems. Physical principles and crystal engineering.}
\newblock 2nd ed. Cham. Switzerland: Springer, 2017{\natexlab{b}}.

\bibitem[Workman and Others(2022)]{PDG}
R.~L. Workman and Others.
\newblock {Review of Particle Physics}.
\newblock \emph{PTEP}, 2022:\penalty0 083C01, 2022.
\newblock \doi{10.1093/ptep/ptac097}.

\bibitem[Collaboration(2017)]{CMS2}
CMS Collaboration.
\newblock Technical proposal for a mip timing detector in the cms experiment phase 2 upgrade.
\newblock \emph{Tech. Rep. CERN-LHCC-2017-027. LHCC-P-009}, 2017.

\bibitem[Gundacker et~al.(2021{\natexlab{b}})Gundacker, Pots, Nepomnyashchikh, Radzhabov, Shendrik, Omelkov, Kirm, Acerbi, Capasso, Paternoster, Mazzi, Gola, Chen, and Auffray]{Gundacker_2021}
S~Gundacker, R~H Pots, A~Nepomnyashchikh, E~Radzhabov, R~Shendrik, S~Omelkov, M~Kirm, F~Acerbi, M~Capasso, G~Paternoster, A~Mazzi, A~Gola, J~Chen, and E~Auffray.
\newblock Vacuum ultraviolet silicon photomultipliers applied to baf2 cross-luminescence detection for high-rate ultrafast timing applications.
\newblock \emph{Physics in Medicine \& Biology}, 66\penalty0 (11):\penalty0 114002, jun 2021{\natexlab{b}}.
\newblock \doi{10.1088/1361-6560/abf476}.
\newblock URL \url{https://dx.doi.org/10.1088/1361-6560/abf476}.

\bibitem[Pagano et~al.(2022)Pagano, Kratochwil, Frank, Gundacker, Paganoni, Pizzichemi, Salomoni, and Auffray]{pagano_2022}
Fiammetta Pagano, Nicolaus Kratochwil, Isabel Frank, Stefan Gundacker, Marco Paganoni, Marco Pizzichemi, Matteo Salomoni, and Etiennette Auffray.
\newblock A new method to characterize low stopping power and ultra-fast scintillators using pulsed x-rays.
\newblock \emph{Frontiers in Physics}, 10, 2022.
\newblock ISSN 2296-424X.
\newblock \doi{10.3389/fphy.2022.1021787}.
\newblock URL \url{https://www.frontiersin.org/articles/10.3389/fphy.2022.1021787}.

\bibitem[\textit{et al.}(2013{\natexlab{b}})]{Gundacker-2013}
S.~Gundacker \textit{et al.}
\newblock Time of flight positron emission tomography towards 100 ps resolution with l(y)so: an experimental and theoretical analysis.
\newblock \emph{J Instrum}, 8\penalty0 (P07014), 2013{\natexlab{b}}.

\bibitem[Alenkov et~al.(2019)Alenkov, Buzanov, Dosovitskiy, Egorychev, Fedorov, Golutvin, Guz, Jacobsson, Korjik, Kozlov, Mechinsky, Schopper, Semennikov, Shatalov, and Shmanin]{ALENKOV_2019}
V.~Alenkov, O.~Buzanov, G.~Dosovitskiy, V.~Egorychev, A.~Fedorov, A.~Golutvin, Yu. Guz, R.~Jacobsson, M.~Korjik, D.~Kozlov, V.~Mechinsky, A.~Schopper, A.~Semennikov, P.~Shatalov, and E.~Shmanin.
\newblock Irradiation studies of a multi-doped gd3al2ga3o12 scintillator.
\newblock \emph{Nuclear Instruments and Methods in Physics Research Section A: Accelerators, Spectrometers, Detectors and Associated Equipment}, 916:\penalty0 226--229, 2019.
\newblock ISSN 0168-9002.
\newblock \doi{https://doi.org/10.1016/j.nima.2018.11.101}.
\newblock URL \url{https://www.sciencedirect.com/science/article/pii/S0168900218317480}.

\bibitem[Lucchini et~al.(2016)Lucchini, Pauwels, Blazek, Ochesanu, and Auffray]{Lucchini_2016}
M.~T. Lucchini, K.~Pauwels, K.~Blazek, S.~Ochesanu, and E.~Auffray.
\newblock Radiation tolerance of luag:ce and yag:ce crystals under high levels of gamma- and proton-irradiation.
\newblock \emph{IEEE Transactions on Nuclear Science}, 63\penalty0 (2):\penalty0 586--590, 2016.
\newblock \doi{10.1109/TNS.2015.2493347}.

\bibitem[Hu et~al.(2018)Hu, Yang, Zhang, Zhu, Kapustinsky, Nelson, and Wang]{Zhu_2018}
Chen Hu, Fan Yang, Liyuan Zhang, Ren-Yuan Zhu, Jon Kapustinsky, Ron Nelson, and Zhehui Wang.
\newblock Proton-induced radiation damage in baf2, lyso, and pwo crystal scintillators.
\newblock \emph{IEEE Transactions on Nuclear Science}, 65\penalty0 (4):\penalty0 1018--1024, 2018.
\newblock \doi{10.1109/TNS.2018.2808841}.

\end{thebibliography}
\end{document}